\documentstyle[epsfig,12pt]{article}
\def\beqn{\begin{eqnarray}}
\def\eeqn{\end{eqnarray}}

\def\beq{\begin{equation}}
\def\eeq{\end{equation}}
\def\ba{\beq\new\begin{array}{c}}
\def\ea{\end{array}\eeq}

\newcommand{\nfour}{${\cal N}=4\;$}
\newcommand{\ntwo}{${\cal N}=2\;$}
\newcommand{\none}{${\cal N}=1\;$}

\newcommand{\pt}{\partial}

\newcommand{\qt}{\tilde q}

\begin{document}


\begin{titlepage}

\begin{flushright}
FTPI-MINN-04/59, UMN-TH-2336/04\\
ITEP-TH-01/05\\
January 26, 2005
\end{flushright}

\begin{center}

{\Large \bf    Non-Abelian Flux Tubes in
\boldmath{\none} SQCD: \\[1mm] Supersizing
World-Sheet Supersymmetry }
\end{center}

\begin{center}
{\bf M.~Shifman$^{a,b}$ and \bf A.~Yung$^{a,b,c}$}
\end {center}
\vspace{0.3cm}
\begin{center}

$^a${\it  William I. Fine Theoretical Physics Institute,
University of Minnesota,
Minneapolis, MN 55455, USA}\\
$^{b}${\it Petersburg Nuclear Physics Institute, Gatchina, St. Petersburg
188300, Russia}\\
$^c${\it Institute of Theoretical and Experimental Physics, Moscow
117259, Russia}
\end{center}

\begin{abstract}

We consider non-Abelian 1/2 BPS flux tubes (strings)
in a deformed $\mbox{\ntwo}$ su\-persymmetric gauge theory, with
mass terms $\mu_{1,2}$
of the adjoint fields breaking \ntwo down to \none$\!\!\!.$
The main feature of the non-Abelian strings
is the occurrence of orientational moduli associated with
the possibility of rotations of
their color fluxes inside a global SU($N$) group. The bulk four-dimensional
theory has four supercharges; half-criticality of the non-Abelian
strings would imply then $\mbox{\none}$ supersymmetry on the world sheet,
i.e. two supercharges. In fact, superalgebra of the reduced moduli
space has {\em four} supercharges. Internal dynamics of
the  orientational moduli are described by two-dimensional
$CP(N-1)$ model on the string world sheet. We focus mainly on
the SU(2) case, i.e. $CP(1)$ world-sheet theory.
We show that non-Abelian BPS strings exist for all values
of $\mu_{1,2}$.  The low-energy theory of moduli
is indeed  $CP(1)$, with four supercharges,
in a wide region of breaking parameters $\mu_{1,2}$.
Only in the limit of very large $\mu_{1,2}$,
above some critical value,
the \ntwo world-sheet supersymmetry breaks down to \none.

We observe ``supersymmetry emergence"
for the flux-tube junction (confined monopole):
the kink--monopole is half-critical considered
from the standpoint of the world-sheet $CP(1)$ model
(i.e. two supercharges conserved),
while in the bulk \none theory there is
no   monopole central charge at all.

\end{abstract}

\end{titlepage}
\tableofcontents

\newpage

\section{Introduction}
\label{intro}
\renewcommand{\theequation}{\thesection.\arabic{equation}}
\setcounter{equation}{0}

Recently, significant progress has been achieved
in obtaining non-Abelian strings in four-dimensional
Yang--Mills theories (to be referred to as {\em bulk theories}), both
$\mbox{\ntwo}$ supersymmetric and non-supersymmetric
\cite{recent,Hanany,Auzzi,ShifmanYung,Tong,HananyTong,Markov,GSY}.
A variety of models emerged which  support non-Abelian
magnetic flux tubes and non-Abelian confined magnetic monopoles
at weak coupling. The non-Abelian strings are characterized by the
presence of orientational moduli associated with the rotation of their
color flux in the non-Abelian gauge group SU($N$).
In supersymmetric bulk theories the non-Abelian strings are
1/2 BPS saturated. The low-energy world-sheet theory
describing moduli dynamics turns out to be
supersymmetric $CP(N-1)$ model.

As well-known, critical solitons in field theory generally exhibit a moduli
space ${\cal M}$ which (locally) admits the decomposition
\beq
{\cal M}\to {\cal M}_{\rm SUSY}\times \widetilde{\cal M}
\label{one}
\eeq
where ${\cal M}_{\rm SUSY}$ refers to the sector associated with
bosonic generators in the superalgebra which are broken by the given soliton,
by virtue of the introduction of central charges \cite{WO}, plus
their fermionic counterparts. In the case at hand, magnetic flux tubes,
two translations are spontaneously broken. Realization of
supersymmetry in this sector, associated with the unbroken generators
(a half of translations and supertranslations are unbroken
in the problem to be considered below) is
fully fixed by flat geometry\,\footnote{See  Sect.~\ref{7} for more precise statements.}.

At the same time, $\widetilde{\cal M}$ in Eq.~(\ref{one}),  the
{\em reduced moduli space}, associated with internal symmetries
and the corresponding moduli, can have realizations of supersymmetry
that are more contrived. A   phenomenon
of this type --- {\em supersymmetry enhancement} --- was discovered
in Ref.~\cite{Adam2} in the domain wall problem.
The world-sheet dynamics on $\widetilde{\cal M}$, at the level
of two derivatives, were described \cite{Adam2} by a three-dimensional
model which had twice more supercharges than one could have {\em a priori}
expected.

In the present work we report similar results for the
non-Abelian strings which emerge as topological defects in some \none
four-dimensional super-Yang--Mills models with matter.
The bulk model has four
supercharges; the strings under consideration are 1/2 BPS.
One could expect two supercharges in the world-sheet algebra.
At the same time, the low-energy theory of moduli
on the string world sheet --- the $CP(N-1)$ model ---
has four supercharges. Two extra
(or ``supernumerary")
supercharges which are realized on
$\widetilde{\cal M}$ can{\em not} be lifted to supercharges of the bulk theory.
Thus, the phenomenon of supersymmetry enhancement, or
{\em supersizing of the
world-sheet supersymmetry},\,\footnote{{\em Supersizing supersymmetry} is a part of the title of th
e talk delivered
by Adam Ritz at {\sl Continuous Advances
in QCD 2004}, see \cite{Adam1}, devoted to supersymmetry enhancement
in the problem of domain walls.} is of a rather general nature
and is not rare.
It has a geometric origin and can be traced back to the
K\"{a}hler structure of the reduced moduli
space.\footnote{That supersymmetry enhancement could take place in flux-tube problems
was conjectured \cite{Adam1} shortly after publication
\cite{Adam2}.}

A particular bulk theory we will deal with is a
deformed $\mbox{\ntwo}$ su\-persymmetric SU($N$)$\times$U(1) theory.
This model has been already heavily exploited \cite{ShifmanYung} in the context of
non-Abelian strings previously. Deformation discussed in
 \cite{ShifmanYung} was a linear in ${\cal A}$ superpotential term, where
${\cal A}$ is the adjoint superfield $\in $U(1). This deformation
is known to be \ntwo preserving. Now, instead, we introduce
mass terms $\mu_{1,2}$
of the adjoint superfields  ${\cal A}^a$ and  ${\cal A}$ which
certainly
break \ntwo down to \none$\! .$

Thus, the bulk four-dimensional
theory has four supercharges. Concentrating mainly on
the simplest case of SU(2)$\times$U(1)
we  construct 1/2 BPS non-Abelian string solution exploiting techniques
worked out previously. Because of
half-criticality of our solution, {\em a priori}
we could expect two supercharges on the reduced moduli
space, i.e. an \none low-energy theory of moduli.
This is not what actually happens. We show, by performing
an explicit analysis
of the zero modes, that the world-sheet
theory on the reduced moduli space is the supersymmetric
$CP(1)$ model (at the level of two derivatives).
This model has \ntwo i.e. four supercharges (for a review see e.g.
\cite{NSVZsigma}).

The (real) dimension of the bosonic part
of $\widetilde{\cal M}$ is two.
The necessary condition for the enhancement of supersymmetry is
the occurrence of four fermion zero modes. Thus, the most crucial
and most technically involved part of the analysis of the zero modes is that of
the fermion zero modes. Their construction is carried out explicitly,
including two extra modes.
Once we obtain four fermion zero modes and introduce corresponding
fermion moduli (four), combining this with the knowledge that
\none supersymmetry on the world sheet is automatic,
the K\"{a}hler structure of $\widetilde{\cal M}$
immediately implies the full-blown \ntwo.

Then we address the issue  of the evolution of the mass deformation.
Indeed, as $\mu_{1,2}\to\infty$ (in fact, we only need
$\mu\gg\sqrt\xi$, where $\xi$ is a Fayet--Iliopoulos parameter),
the adjoint fields ${\cal A}^a$ and  ${\cal A}$
become very heavy and decouple from the bulk theory altogether,
leading to   \none  SQCD with the gauge group SU(2)$\times$U(1).
It is known \cite{GS} that  \none SQCD
admits only Abelian BPS strings. The question is what happens
with our non-Abelian 1/2 BPS strings as the parameters $\mu_{1,2}$ grow.

This question turns out to be subtle.
It turns out that the parameter $\xi/\mu$ plays the role of an
infrared regulator. Physically, at $\mu\gg\sqrt\xi$ the adjoint fields
do decouple. However, in the limit $\mu\to \infty$, after the decoupling,
the emerging
\none  SQCD develops a Higgs branch, which is absent for any finite $\mu$.
At  any finite $\mu$ the vacuum manifold is an isolated point,
which makes the string solution, as well as zero modes, well-defined.
If $\mu$ is large but finite, the mass of the would-be moduli corresponding to
the ``motion" along the Higgs branch is $\sim \xi/\mu$.

Thus, there is a seemingly irreconcilable
contradiction. On the one hand, it is clear
that at $\mu\gg\sqrt\xi$ we {\em must} recover \none SQCD.
On the other hand, in the BPS string analysis
the limit $\mu\to 0$ seemingly can{\em{not}} be taken.

A way out was in fact suggested in the literature in
the context of a similar
problem \cite{Yung:1999du}. In Ref.~\cite{Yung:1999du}
Abrikosov-Nielsen-Olesen (ANO) strings \cite{ANO}
were considered on the Higgs branch of an \ntwo
gauge theory (with massive fundamental matter).
Common wisdom says \cite{PRTT} that there are no ANO strings
in this case (it would be more accurate to say that they
inflate and  become {\em infinitely} thick), because of the same infrared problem.
It was discovered, however, that strings of
finite size $L$ are perfectly well-defined, no matter how large
$L$ is. The role of $L$ is to provide an infrared regularization.
The string thickness was found \cite{Yung:1999du} to be proportional to
$\ln L$, while the string mass $\sim L/\ln L$ rather than
pure $L$  in the classical ANO case.

If we do the same thing in our problem --- i.e. consider a finite-length
string  --- the limit $\mu\to\infty$
will become perfectly
well-defined. The parameter $ \mu /\xi $ will be replaced by $L$,
which will provide infrared regularization. Unlike the problem
considered in \cite{Yung:1999du}, in the present case
the infrared divergence does not appear in the bosonic string
solution {\em per se}. It is only the  ``extra" fermion zero
mode normalization that is
plagued by logarithmic divergence.

There is a price one has to pay for the  finite-length regularization ---
the loss of  ``BPS-ness." Since  ``BPS-ness" is a convenient feature,
we find the finite-$\mu$ regularization to be more appropriate,
even though it requires inclusion of the adjoint fields
in the bulk Lagrangian. This seems to be a smaller price.
Once we stick to the finite-$\mu$ regularization
and normalizability of four fermion zero modes is achieved, the
low-energy theory of moduli exhibits
supersymmetry enhancement. The normalizing parameter,
which depends on $\mu$ logarithmically,
can be absorbed in the definition of the moduli fields
and does not show up explicitly.

Thus, if \none bulk theory has an isolated vacuum
(no Higgs branch) we can state with certainty that
the low-energy moduli theory
on the world sheet of the non-Abelian BPS string
is indeed  $CP(1)$, with four supercharges,
as long as we limit ourselves to two-derivative terms
in the  world-sheet Lagrangian.

It is necessary to stress that although many features
of the analysis reported here are parallel to
those of the domain-wall problem \cite{Adam2}, some important
features are rather different. In
particular, a K\"ahler structure for the moduli space, which appears
automatic, is not sufficient now, generally speaking,
for  enhanced SUSY, since the Lorentz
invariance in 1+1 dimensions imposes no useful constraints
(as opposed to the situation \cite{Adam2} in 1+2 dimensions).
Indeed,  in
the pure \none limit (i.e with $\mu_{1,2} =\infty$), the K\"ahler
structure
for the
bosonic moduli space persists. Then the minimal \none world-sheet
SUSY will be realized in the chiral (0,2) form consistent with the complex
structure.

In the flux-tube problem it is not the Lorentz invariance which ensures the
one-to-two matching of bosonic versus fermionic zero modes, but, instead,
the possibility of embedding the system within \ntwo SQCD. This
possibility was not available in the domain-wall case \cite{Adam2}.

As a warm up exercise we will also consider a seemingly well-studied
problem of the ANO strings in \none SQED. Of course, in this case, the
internal moduli space $\widetilde{\cal M}$ is absent.
However, following the same line of reasoning as in
the case of non-Abelian strings above, we can start from \ntwo SQED
\cite{VY}
(eight supercharges),  construct
the Abelian half-critical string which has four fermion moduli
in ${\cal M}_{\rm SUSY}$ and then make the adjoint mass deformation
term very large effectively returning to \none SQED. For arbitrarily large but
finite $\mu$ we will keep all {\em four} fermion zero modes:
two natural and two ``extra."
Correspondingly, we will keep the
\ntwo theory of moduli from ${\cal M}_{\rm SUSY}$. Of course,
in this methodical example it is a trivial free field theory (in  1+1 dimensions).

\section{ The bulk theory}
\label{bth}
\renewcommand{\theequation}{\thesection.\arabic{equation}}
\setcounter{equation}{0}

In this section we will briefly describe the bulk theories we will   deal
with. $\mbox{ \ntwo}$ SQED is discussed in detail in
Ref.~\cite{VY}
while the version of SQCD we will focus on is thoroughly discussed in
Refs.~\cite{Auzzi,ShifmanYung}.

\subsection{Abelian bulk theory}
\label{abt}

Let us denote scalar and fermion fields in
the ``quark'' hypermultiplets    as
$q$, $\tilde{q}$ and $\psi$, and $\tilde{\psi}$, respectively.
Note that the scalars form a doublet under the action of global  SU(2)$_R$
group,
$q^f=(q,\bar{\tilde{q}})$.
In terms of these fields the action of \ntwo SQED deformed
by the (\ntwo)-breaking mass term $\mu$
of the adjoint field $a$
takes the
form\,\footnote{Here and below we use a
formally  Euclidean notation, e.g.
$F_{\mu\nu}^2 = 2F_{0i}^2 + F_{ij}^2$,
$\, (\partial_\mu a)^2 = (\partial_0 a)^2 +(\partial_i a)^2$, etc.
This is appropriate since we are
going to study static (time-independent)
field configurations, and $A_0 =0$. Then the Euclidean action is
nothing but the energy functional. Furthermore, we
define $\sigma^{\alpha\dot{\alpha}}=(1,-i\vec{\tau})$,
$\bar{\sigma}_{\dot{\alpha}\alpha}=(1,i\vec{\tau})$. Lowing and raising
of spinor indices
is performed by
virtue of the antisymmetric tensor defined as
$\varepsilon_{12}=\varepsilon_{\dot{1}\dot{2}}=1$,
$\varepsilon^{12}=\varepsilon^{\dot{1}\dot{2}}=-1$.
The same raising and lowering convention applies to the flavor SU(2)
indices $f$, $g$, etc.,  see  \cite{ShifmanYung}. }
\begin{eqnarray}
S_{\rm SQED} &=&\int d^4 x \left\{ \frac1{4 e^2}F_{\mu\nu}^2 +
\frac1{e^2} \left|\partial_{\mu}a\right|^2
+\bar{\nabla}_{\mu}\bar{q}\nabla_{\mu}q+
\bar{\nabla}_{\mu}\tilde{q}\nabla_{\mu}\bar{\tilde{q}} \right.
\nonumber\\[3mm]
&+&\frac{e^2}{8}\left( |q|^2 -|\tilde{q}|^2-2\xi \right)^2 +
\frac{e^2}{2}\left|\qt q +\sqrt{2}\mu a\right|^2
\nonumber\\[3mm]
&+&\left.\frac12 \left( |q|^2+|\tilde{q}|^2\right)\, \left| a \right|^2
\right\}+\mbox{fermion part}\,,
\label{qed}
\end{eqnarray}
where
$$
\nabla_{\mu}=\partial_{\mu} -\frac{i}{2}A_{\mu}\,,\qquad
\bar{\nabla}_{\mu}=\partial_{\mu} +\frac{i}{2}A_{\mu}\,,
$$
while  $\xi$ is the Fayet--Iliopoulos  (FI) parameter.
The vacuum in this theory is determined (up to gauge transformations)
by the following vacuum expectation values (VEV's):
\beq
\langle q\rangle =\sqrt{\xi},\,\,\, \langle \tilde{q}\rangle =0,\,\,\,
\langle  a\rangle =0.
\label{qedvac}
\eeq
The nonvanishing VEV of the squark field breaks U(1) gauge symmetry giving mass
to the photon.

The mass spectrum of the theory in the vacuum (\ref{qedvac}) was studied
in Ref.~\cite{VY}, see also \cite{EY}.
At non-zero $\mu$, extended \ntwo supersymmetry
in (\ref{qed}) is broken down to \none and the states come in \none
supermultiplets. The massive vector multiplet has the mass
\beq
m_{\gamma}=\frac{g}{\sqrt{2}}\,\sqrt{\xi},
\label{abphmass}
\eeq
while two chiral multiplets acquire masses
\beq
m^{\pm}=\frac{g}{\sqrt{2}}\,\sqrt{\xi\lambda^{\pm}},
\label{abmass}
\eeq
where $\lambda{\pm}$ are two roots of the quadratic equation
\beq
\lambda^2-\lambda(2+\omega^2) +1=0
\label{abquadeq}
\eeq
and $\omega$ is the \ntwo breaking parameter
\beq
\omega=\frac{\sqrt{2}g\mu}{\sqrt{\xi}}\,.
\label{abomega}
\eeq

At $\mu=0$ one gets
$$\lambda^{\pm}=1\,,$$
and all states listed above form
the bosonic part of one
long  \ntwo massive vector multiplet \cite{VY}.
As we switch the parameter
$\mu$ on, this \ntwo vector multiplet splits into one vector and two chiral
multiplets of \none supersymmetric theory.

In the limit of $\mu\to \infty$ the heavy neutral field $a$ and its superpartners
can be integrated out \cite{KSS,GVY,VY} leading to \none SQED
\beq
S=\int d^4x \left\{\frac1{4e^2}
\left(F_{\mu\nu}\right)^2
+ \left|\nabla_{\mu}
q\right|^2 + \left|\nabla_{\mu} \bar{\tilde{q}}\right|^2
+
\frac{e^2}{8}
\left(\bar{q} q - \tilde{q} \bar{\tilde{q}}-\xi\right)^2
\right\}\,.
\label{noneqed}
\eeq
This theory has a two-dimensional Higgs branch of a hyperbolic form.
As we increase $\mu$  in
(\ref{qed}) we arrive, in the limit $\mu\to \infty$,
at a base point on this Higgs branch with
$\langle \tilde{q}\rangle =0$.

\subsection{Non-Abelian bulk theory}
\label{nabt}

The content of this section  is a direct
non-Abelian generalization of Sect.~\ref{abt}.
The gauge symmetry of the model we will use   is SU(2)$\times$U(1).
Besides the gauge bosons, gauginos and their  superpartners,
it has  a matter sector consisting of two ``quark" hypermultiplets,
with  {\em degenerate}  masses. In addition, we
introduce a Fayet--Iliopoulos $D$-term for the U(1) gauge field
which triggers the quark condensation.

Let us first discuss the undeformed theory with
 \ntwo. The  superpotential has the form
 \beq
{\cal W}_{{\cal N}=2} =\frac{1}{\sqrt 2} \sum_{A=1}^2
\left( \tilde q_A {\cal A}
q^A +  \tilde q_A {\cal A}^a\,\tau^a  q^A\right)
\label{superpot}
\eeq
where ${\cal A}^a$ and ${\cal A}$ are  chiral superfields, the ${\cal N}=2$
superpartners of the gauge bosons of SU(2)  and  U(1), respectively.
{}Furthermore, $q_A$ and $\tilde q_A$ ($A=1,2$) represent two
matter hypermultiplets. The flavor index is denoted by $A$. Thus,
in our model the number of colors equals the number of flavors.

Next we add a superpotential mass term which breaks
supersymmetry down to \none, namely,
\beq
{\cal W}_{{\cal N}=1}=\frac{\mu_1}{2} {\cal A}^2
+ \frac{\mu_2}{2} ({\cal A}^a)^2,
\label{superpotbr}
\eeq
where $\mu_1$ and $\mu_2$ are mass parameters for the chiral
superfields in \ntwo gauge supermultiplets,
U(1) and SU(2) respectively. Clearly, the mass term
(\ref{superpotbr}) splits these supermultiplets, breaking
\ntwo supersymmetry down to \none.

The bosonic part of our SU(2)$\times$U(1)
theory has  the form
\beqn
S&=&\int d^4x \left[\frac1{4g^2_2}
\left(F^{a}_{\mu\nu}\right)^2 +
\frac1{4g^2_1}\left(F_{\mu\nu}\right)^2
+
\frac1{g^2_2}\left|D_{\mu}a^a\right|^2 +\frac1{g^2_1}
\left|\partial_{\mu}a\right|^2 \right.
\nonumber\\[4mm]
&+&\left. \left|\nabla_{\mu}
q^{A}\right|^2 + \left|\nabla_{\mu} \bar{\tilde{q}}^{A}\right|^2
+V(q^A,\tilde{q}_A,a^a,a)\right]\,.
\label{model}
\eeqn
Here $D_{\mu}$ is the covariant derivative in the adjoint representation
of  SU(2),
while
\beq
\nabla_\mu=\partial_\mu -\frac{i}{2}\; A_{\mu}
-i A^{a}_{\mu}\, \frac{\tau^a}{2},
\label{defnabla}
\eeq
and $\tau^a$ are the SU(2) Pauli matrices. The coupling constants $g_1$ and $g_2$
correspond to the U(1)  and  SU(2)  sectors, respectively.
With our conventions the U(1) charges of the fundamental matter fields are $\pm 1/2$.

\vspace{1mm}

The potential $V(q^A,\tilde{q}_A,a^a,a)$ in the Lagrangian (\ref{model})
is a sum of  various $D$ and  $F$  terms,
\beqn
V(q^A,\tilde{q}_A,a^a,a) &=&
 \frac{g^2_2}{2}
\left( \frac{1}{g^2_2}\,  \varepsilon^{abc} \bar a^b a^c
 +
 \bar{q}_A\,\frac{\tau^a}{2} q^A -
\tilde{q}_A \frac{\tau^a}{2}\,\bar{\tilde{q}}^A\right)^2
\nonumber\\[3mm]
&+& \frac{g^2_1}{8}
\left(\bar{q}_A q^A - \tilde{q}_A \bar{\tilde{q}}^A-2\xi\right)^2
\nonumber\\[3mm]
&+& \frac{g^2_2}{2}\left| \tilde{q}_A\tau^a q^A +\sqrt{2}\mu_2a^a\right|^2+
\frac{g^2_1}{2}\left| \tilde{q}_A q^A +\sqrt{2}\mu_1 a \right|^2
\nonumber\\[3mm]
&+&\frac12\sum_{A=1}^2 \left\{ \left|(a +\tau^a a^a)q^A
\right|^2 +
\left|(a +\tau^a a^a)\bar{\tilde{q}}_A
\right|^2 \right\}\,,
\label{pot}
\eeqn
where the sum over repeated flavor indices $A$ is implied.
The first and second lines here represent   $D$   terms, the third line
the $F_{\cal A}$ terms,
while the fourth line represents the squark $F$ terms.
We also introduced the Fayet--Iliopoulos  $D$-term for the U(1) field,
with the FI parameter $\xi$ in (\ref{pot}),
much in the same way as in Sect.~\ref{abt}.
Note that the Fayet--Iliopoulos term does not
break \ntwo supersymmetry \cite{matt,VY}. The parameters which do
break  \ntwo   down to \none are $\mu_1$ and $\mu_2$.

The Fayet--Iliopoulos term triggers the spontaneous breaking
of the gauge symmetry. The vacuum expectation values (VEV's)
of the squark fields can be chosen as
\beqn
\langle q^{kA}\rangle &=&\sqrt{
\xi}\, \left(
\begin{array}{cc}
1 & 0 \\
0 & 1\\
\end{array}
\right),\,\,\,\langle \bar{\tilde{q}}^{kA}\rangle =0,
\nonumber\\[3mm]
k&=&1,2,\qquad A=1,2\,,
\label{qvev}
\eeqn
{\em up to gauge rotations},
while the VEV's of adjoint fields are given by
\beq
\langle a^a\rangle =0,\,\,\,\,\langle a\rangle =0.
\label{avev}
\eeq
Here we write down  $q$ as a $2\times 2$ matrix, the first
superscript ($k=1,2$) refers to SU(2) color, while the second
($A=1,2$) to flavor.

The color-flavor locked form of the quark VEV's in
Eq.~(\ref{qvev}) and the absence of VEV of the adjoint scalar $a^a$ in
Eq.~(\ref{avev})
results in the fact that, while the theory is fully Higgsed, a diagonal
SU(2)$_{C+F}$ survives as a global symmetry.
This is a particular case  of the Bardakci-Halpern mechanism \cite{BarH}.
The presence of this symmetry leads to the emergence of
orientational zero modes of $Z_2$ strings in the model (\ref{model})
\cite{Auzzi}.

Note that VEV's (\ref{qvev}) and  (\ref{avev})  do not depend on
the supersymmetry breaking parameters $\mu_1$ and $\mu_2$. This
is because our choice of parameters in (\ref{model}) ensures
vanishing of the adjoint VEV's, see (\ref{avev}). In particular, we have
the same pattern of symmetry breaking all the way up to very large
$\mu_1$ and $\mu_2$, where the adjoint fields decouple.

With two matter hypermultiplets, the  SU(2) part of the gauge group
is asymptotically free,  implying generation of a dynamical scale
 $\Lambda$.
If descent to  $\Lambda$ were uninterrupted, the gauge coupling
$g_2^2$ would explode at this scale.
Moreover,  strong coupling effects in the SU(2) subsector at the
scale $\Lambda$ would break the  SU(2) subgroup through the
Seiberg-Witten mechanism \cite{SW}.  Since we want to stay
at weak coupling   we assume
that $\sqrt{\xi}\gg \Lambda$,
so that the SU(2) coupling running is frozen by the squark condensation
at a small value, namely,
\beq
\frac{8\pi^2}{g_2^2}=2\ln{\frac{\sqrt{\xi}}{\Lambda}} +\cdots \gg 1\,.
\label{g2}
\eeq

Now let us discuss the mass spectrum in the theory (\ref{model}). Since
both U(1) and SU(2) gauge groups are broken by the squark condensation,
all gauge bosons become massive. From (\ref{model}) we get for the U(1)
gauge boson
\beq
m_{{\rm U}(1)}=g_1\sqrt{\xi}\,,
\label{phmass}
\eeq
while   three gauge bosons of the SU(2) group acquire the same mass
\beq
m_{{\rm SU}(2)}=g_2\sqrt{\xi}\,.
\label{wmass}
\eeq

To get the masses of the scalar bosons we expand the potential (\ref{pot})
near the vacuum (\ref{qvev}), (\ref{avev}) and diagonalize the
corresponding mass matrix. The four components of the
eight-component\,\footnote{We mean here eight {\em real} components.}
scalar $q^{kA}$
are eaten by the Higgs mechanism for U(1) and SU(2)
gauge groups. Another four components are split as follows:
one component acquires the mass (\ref{phmass}). It becomes
 a scalar component of  a massive \none vector U(1) gauge multiplet.
Other three components acquire masses (\ref{wmass}) and become
scalar superpartners of the SU(2) gauge boson in \none massive gauge
supermultiplet.

Other 16 real scalar components of fields $\tilde{q}_{Ak}$, $a^a$ and $a$
produce the following states: two states acquire mass
\beq
m_{{\rm U}(1)}^{+}=g_1\sqrt{\xi\lambda_1^{+}}\,,
\label{u1m1}
\eeq
while the mass of other two states is given by
\beq
m_{{\rm U}(1)}^{-}=g_1\sqrt{\xi\lambda_1^{-}}\,,
\label{u1m2}
\eeq
where $\lambda_1^{\pm}$ are two roots of the quadratic equation
\beq
\lambda_i^2-\lambda_i(2+\omega^2_i) +1=0\,,
\label{queq}
\eeq
for $i=1$. Here we introduced two \ntwo supersymmetry breaking
parameters associated with U(1) and SU(2) gauge groups, respectively,
\beq
\omega_1=\frac{g_1\mu_1}{\sqrt{\xi}}\, ,\qquad
\omega_2=\frac{g_2\mu_2}{\sqrt{\xi}}\,.
\label{omega}
\eeq
Furthermore,
other 2$\times$3=6 states acquire mass
\beq
m_{{\rm SU}(2)}^{+}=g_2\sqrt{\xi\lambda_2^{+}}\,,
\label{su2m1}
\eeq
while the rest 2$\times$3=6 states also become massive, their mass is
\beq
m_{{\rm SU}(2)}^{-}=g_2\sqrt{\xi\lambda_2^{-}}\,.
\label{su2m2}
\eeq
Here $\lambda_2^{\pm}$ are two roots of the quadratic equation
(\ref{queq}) for $i=2$. Note that all states come either as  singlets
or triplets of unbroken SU(2)$_{C+F}$.

When the supersymmetry breaking parameters $\omega_{i}$ vanish, the
masses (\ref{u1m1}) and (\ref{u1m2}) coincide with the U(1) gauge
boson mass (\ref{phmass}). The corresponding states form bosonic part of \ntwo
long massive U(1) vector supermultiplet \cite{VY}.  With non-zero
$\omega_1$ this supermultiplet splits into massive \none vector multiplet
with mass (\ref{phmass}), and two chiral multiplets with masses
(\ref{u1m1}) and (\ref{u1m2}). The same happens to states with masses
(\ref{su2m1}) and (\ref{su2m2}).  If $\omega$'s vanish they combine
into the bosonic parts of three  \ntwo massive vector supermultiplets,
with mass
(\ref{wmass}). At non-zero $\omega$'s these multiplets split to three
 \none vector multiplets (for SU(2) group) with mass (\ref{wmass})
and 2$\times$3 chiral multiplets with masses (\ref{su2m1}) and (\ref{su2m2}).
Note that essentially the same pattern of splitting was found in \cite{VY} for
the Abelian  case, see Sect.~\ref{abt}.

Now let us take a closer look at the spectrum obtained above in the limit
of large \ntwo supersymmetry breaking parameters $\omega_i$, $$\omega_i\gg 1\,.$$
In this limit
the larger masses $m_{{\rm U}(1)}^{+}$ and $m_{{\rm SU}(2)}^{+}$ become
\beq
m_{{\rm U}(1)}^{+}= m_{{\rm U}(1)}\omega_1=g_1^2\mu_1\,,\qquad
m_{{\rm SU}(2)}^{+}= m_{{\rm SU}(2)}\omega_2=g_2^2\mu_2\, .
\label{amass}
\eeq
Clearly, in the limit $\mu_i\to \infty$ these are
the masses of the heavy adjoint
scalars $a$ and $a^a$. At $\omega_i\gg 1$ these fields decouple and
can be integrated out.

The low-energy bulk theory in this limit
contains massive gauge  \none multiplets and chiral multiplets with
lower masses $m^{-}$. Equation (\ref{queq}) gives for these masses
\beq
m_{{\rm U}(1)}^{-}= \frac{m_{{\rm U}(1)}}{\omega_1}
=
\frac{\xi}{\mu_1}\,,\qquad
m_{{\rm SU}(2)}^{-}= \frac{m_{{\rm SU}(2)}}{\omega_2}
=
\frac{\xi}{\mu_2}\,.
\label{light}
\eeq
In  the limit of infinite $\mu_i$ these masses tend to zero.
This fact reflects the emergence of a Higgs branch in \none SQCD,
see also Eq.~(\ref{noneqed}). To
observe the Higgs branch it is instructive to inspect
the transition to $\mu =\infty$ in (\ref{model}). Equation
(\ref{model}) flows to \none
SQCD with the gauge group SU(2)$\times$U(1) and the
Fayet--Iliopoulos $D$-term,
\beqn
S&=&\int d^4x \left\{\frac1{4g^2_2}
\left(F^{a}_{\mu\nu}\right)^2 +
\frac1{4g^2_1}\left(F_{\mu\nu}\right)^2
+ \left|\nabla_{\mu}
q^{A}\right|^2 + \left|\nabla_{\mu} \bar{\tilde{q}}^{A}\right|^2
\right.
\nonumber\\[4mm]
& + &
\left.
\frac{g^2_2}{2}\left(
\bar{q}_A\,\frac{\tau^a}{2} q^A -
\tilde{q}_A \frac{\tau^a}{2}\,\bar{\tilde{q}}^A\right)^2
+ \frac{g^2_1}{8}
\left(\bar{q}_A q^A - \tilde{q}_A \bar{\tilde{q}}^A-2\xi\right)^2
\right\}\,.
\label{noneqcd}
\eeqn
All $F$ terms disappear in this limit and we are left only with $D$ terms.
{}For 16 real components of $q$ and $\tilde{q}$
we have four $D$-term constraints in (\ref{noneqcd}).
Another four phases are
eaten by the Higgs mechanism. Thus, the dimension of the Higgs branch in
(\ref{noneqcd}) is $16-4-4=8$. It can be described in terms of a gauge
invariant meson matrix
\beq
M_A^B=\tilde{q}_A q^B
\label{meson}
\eeq
plus baryon operators\,\footnote{The baryon operators are not
U(1) gauge invariant in the SU(2)$\times$U(1) theory. Their product is
gauge invariant, however.}
\beq
B^{AB}=\frac12 \varepsilon_{kl}q^{kA}q^{lB}\,, \qquad
\tilde{B}_{AB}=\frac12 \varepsilon^{kl}\tilde{q}_{Ak}\tilde{q}_{Bl}\,,
\label{baryons}
\eeq
see \cite{SeibIntril} for a review. These operators are subject to
a classical constraint
\beq
{\rm det}\,  M-\tilde{B}_{AB}B^{AB}=0
\label{classconstr}
\eeq
which gets modified by instanton effects and becomes
\beq
{\rm det}\, M-\tilde{B}_{AB}B^{AB}=\Lambda_{{\cal N}=1}^4
\label{quantconstr}
\eeq
in the quantum theory \cite{SeibIntril}. Here $\Lambda_{{\cal N}=1}$
is the scale of \none SQCD   in terms of the scale $\Lambda$
of the deformed \ntwo theory (\ref{model}); $\Lambda_{{\cal N}=1}$
has the form
\beq
\Lambda_{{\cal N}=1}^4=\mu_2^2\,\Lambda^2.
\label{lambdanone}
\eeq
In order to keep the bulk theory in the
weak coupling regime, in the limit
of large $\mu_i$ we assume that
\beq
\sqrt{\xi}\gg \Lambda_{{\cal N}=1}\,.
\label{weakcoupling}
\eeq
Note that the presence of
the FI term cannot modify (\ref{quantconstr}) because $\xi$ is not
a holomorphic parameter.

The vacuum (\ref{qvev}) corresponds to the base point of this
Higgs branch with $\tilde{q}=0$. In other words, flowing from \ntwo
theory (\ref{model}) we do not recover the whole Higgs branch
of \none SQCD (\ref{noneqcd}). Instead, we arrive only at  an isolated
vacuum,  a base point of the Higgs branch, no matter how large $\mu$ is.

What else is there to say? A question to be discussed is as follows:
how  our solution in which $\tilde q = 0$
can be compatible with the quantum constraint (\ref{quantconstr})?
It seems apparent that
the classical vacuum with $\tilde{q}=0$ at the base of the Higgs
branch no longer exists at the quantum level.

Our analysis is quasiclassical. We start with $\tilde{q}=0$, so
that the corresponding light
moduli are not excited.   Next we
consider quantum corrections. What enters in the constraint
(\ref{quantconstr}) is
the {\em quantum} average of the {\em composite} operator
$\langle \tilde{q} q \rangle$.
The above VEV does not factorize, and
Eq.~(\ref{quantconstr}) can still hold in our solution.
In fact, we expect it to hold. While the light modes fluctuate along
the Higgs branch, the massive modes
fluctuate in the ``orthogonal" directions. Account of these latter fluctuations must
modify the classical constraint (\ref{classconstr})
transforming it into (\ref{quantconstr}).

Certainly, it would be instructive
to check this explicitly. We leave this exercise for future studies.
This issue is of a conceptual importance. Practically, though, it is rather
unimportant since we work in the
regime (\ref {weakcoupling}), so that the quantum deformation
is parametrically small.

\section{Non-Abelian strings}
\label{3}
\renewcommand{\theequation}{\thesection.\arabic{equation}}
\setcounter{equation}{0}

Recently, non-Abelian strings were shown to emerge at weak coupling
\cite{Auzzi,ShifmanYung,HananyTong,Markov} in
\ntwo and deformed \nfour supersymmetric gauge theories
(similar results in three dimensions were obtained in \cite{Hanany}).
The  main feature of  the non-Abelian strings is the
presence of orientational zero modes associated with rotation of their
color flux in the non-Abelian gauge group, which makes such  strings
genuinely non-Abelian. As soon as the solution for
the non-Abelian string
suggested in \cite{Auzzi,ShifmanYung} for \ntwo SQCD does not
depend on the
adjoint fields it can be easily generalized to our model (\ref{model})
with the broken \ntwo supersymmetry. We will carry out this program in  Sect.~\ref{31}.

\subsection{The non-Abelian string solution}
\label{31}

Here we generalize the string solutions found in \cite{Auzzi,ShifmanYung}
to the model (\ref{model}).
Since this model  includes a spontaneously broken gauge U(1),
it supports
conventional Abrikosov-Nielsen-Olesen (ANO) strings \cite{ANO}
in which one can discard the SU($2$)$_{\rm gauge}$ part
of the action.
The topological stability of the ANO string is due to the fact that
$\pi_1({\rm U(1)}) = Z$. These are not the strings we are interested in.
At first sight the triviality of the homotopy group, $\pi_1
({\rm SU}(2)) =0$, implies that there are no other topologically stable
strings.
This impression is false. One can
combine the $Z_2$ center of SU($2$) with the elements
$\exp (i\pi  )\in$U(1) to get a topologically stable string solution
possessing both windings, in SU($2$) and U(1). In other words,
\beq
\pi_1 \left({\rm SU}(2)\times {\rm U}(1)/ Z_2
\right)\neq 0\,.
\eeq
It is easy to see that this non-trivial topology amounts to winding
of just one element of matrix $q_{\rm vac}$, say, $q^{11}$, or
$q^{22}$,  for instance,\footnote{As explained below,
$\alpha$ is the angle of
the coordinate  $\vec{x}_\perp$ in the perpendicular plane.}
\beq
q_{\rm string} = \sqrt{\xi}\,
\left(
\begin{array}{cc}
e^{ i \, \alpha  } & 0  \\
0 &  1 \\
\end{array}\right)\,,
\quad x\to\infty \,.
\label{ansa}
\eeq
Such strings can be called elementary;
their tension is $1/2$ of that of the ANO string.
The ANO string can be viewed as a bound state of
two elementary strings.

More concretely,  the $Z_2$ string solution
(a progenitor of the non-Abelian string) can be written as
follows \cite{Auzzi}:
\beqn
q(x)
&=&
\left(
\begin{array}{cc}
e^{ i \, \alpha  }\phi_1(r) & 0  \\
0 &  \phi_2(r) \\
\end{array}\right),
\nonumber\\[4mm]
A^3_{i}(x)
&=&
 -\varepsilon_{ij}\,\frac{x_j}{r^2}\
\left(1-f_3(r)\right),\;
\nonumber\\[4mm]
A_{i}(x)
&=&
- \varepsilon_{ij}\,\frac{x_j}{r^2}\
\left(1-f(r)\right)\,
\label{znstr}
\eeqn
where $i=1,2$ labels coordinates in the plane orthogonal to the string
axis and $r$ and $\alpha$ are the polar coordinates in this plane. The profile
functions $\phi_1(r)$ and  $\phi_2(r)$ determine the profiles of
the scalar fields,
while $f_{3}(r)$ and $f(r)$ determine the SU($2$) and U(1)
gauge fields of the
string solutions, respectively. These functions satisfy the following
first order equations \cite{Auzzi}:
\beqn
&&
r\frac{d}{{d}r}\,\phi_1 (r)- \frac12\left( f(r)
+  f_3(r) \right)\phi_1 (r) = 0\, ,
\nonumber\\[4mm]
&&
r\frac{d}{{ d}r}\,\phi_2 (r)- \frac12\left(f(r)
-  f_3(r)\right)\phi_2 (r) = 0\, ,
\nonumber\\[4mm]
&&
-\frac1r\,\frac{ d}{{ d}r} f(r)+\frac{g^2_1}{2}\,
\left[\left(\phi_1(r)\right)^2 +\left(\phi_2(r)\right)^2-2\xi\right] =
0\, ,
\nonumber\\[4mm]
&&
-\frac1r\,\frac{d}{{ d}r} f_3(r)+\frac{g^2_2}{2}\,
\left[\left(\phi_1(r)\right)^2 -\left(\phi_2(r)\right)^2\right]  = 0
\, .
\label{foe}
\eeqn
Furthermore, one needs to specify the boundary conditions
which would determine the profile functions in these equations. Namely,
\beqn
&&
f_3(0) = 1\, ,\qquad f(0)=1\, ;
\nonumber\\[4mm]
&&
f_3(\infty)=0\, , \qquad   f(\infty) = 0
\label{fbc}
\eeqn
for the gauge fields, while the boundary conditions for  the
squark fields are
\beqn
\phi_1 (\infty)=\sqrt{\xi}\,,\qquad   \phi_2 (\infty)=\sqrt{\xi}\,,
\qquad \phi_1 (0)=0\, .
\label{phibc}
\eeqn
Note that since the field $ \phi_2 $ does not wind, it need not vanish
at the origin, and, in fact, it does not. Numerical solutions of the
Bogomolny equations (\ref{foe}) for $Z_2$ strings were
found in Ref.~\cite{Auzzi}, see e.g. Figs. 1 and 2 in this paper.

The tension of this elementary string is
\beq
T_1=2\pi\,\xi\, ,
\label{ten}
\eeq
to be compared with  the tension of the ANO
string,
\beq
T_{\rm ANO}=4\pi\,\xi
\label{tenANO}
\eeq
in our normalization.

The elementary strings are {\em bona fide} non-Abelian.
This means that, besides trivial translational
moduli, they give rise to moduli corresponding to spontaneous
breaking of a non-Abelian symmetry. Indeed, while the ``flat"
vacuum (\ref{qvev})
is SU($2$)$_{C+F}$ symmetric, the solution (\ref{znstr})
breaks this symmetry
down to U(1).
This means that the world-sheet (two-dimensional) theory of
the elementary string moduli
is the SU($2$)/U(1) sigma model.
This is also known as $CP(1)$ model.

To obtain the non-Abelian string solution from the $Z_2$ string
(\ref{znstr}) we apply the diagonal color-flavor rotation  preserving
the vacuum (\ref{qvev}). To this end
it is convenient to pass to the singular gauge where the scalar fields have
no winding at infinity, while the string flux comes from the vicinity of
the origin. In this gauge we have
\begin{eqnarray}
q
&=&
U \left(
\begin{array}{cc}
\phi_1(r) & 0  \\[2mm]
0 &  \phi_2(r)
\end{array}\right)U^{-1}\, ,
\nonumber \\[4mm]
A^a_{i}(x)
&=&
n^a \,\varepsilon_{ij}\, \frac{x_j}{r^2}\,
f_3(r)\, ,
\nonumber \\[4mm]
A_{i}(x)
&=&
 \varepsilon_{ij} \, \frac{x_j}{r^2}\,
f(r)\, ,
\label{sna}
\end{eqnarray}
where $U$ is a matrix $\in {\rm SU}(2)$ and
$ n^a$ is a moduli vector
defined as
\beq
n^a\tau^a=U \tau^3 U^{-1},\;\;a=1,2,3.
\label{n}
\eeq
It is  subject to the constraint
\beq
{\vec n}^{\,2} = 1\,.
\label{ensq}
\eeq
At $n=\{0,0, 1\}$ we get the field configuration quoted
in Eq.~(\ref{znstr}).

The vector $n^a$ parametrizes
orientational zero modes of the string associated with flux rotation
in  SU($2$). The presence of these modes makes the string genuinely
non-Abelian. We stress that
the orientational moduli encoded in the vector $n^a$, first
observed in \cite{Hanany, Auzzi}, are {\it not} gauge artifacts.

\subsection{World-sheet effective theory}
\label{32}

In this subsection we
briefly review derivation of the effective world-sheet
theory for the orientational collective coordinates $n^a$ of the non-Abelian
string.  We follow
Ref.~\cite{Auzzi,ShifmanYung}.
(Generalization to the case of SU(N)$\times$U(1) gauge group is done in \cite{GSY}.)
As was already
mentioned, this macroscopic theory is $CP(1)$ model ($CP(N-1)$ model for
the general case of SU(N)$\times$U(1) gauge group)
\cite{Hanany,Auzzi,ShifmanYung,HananyTong,GSY}.

Assume  that the orientational collective coordinates $n^a$
are slowly varying functions of the string world-sheet coordinates
$x_k$, $k=0,3$. Then the moduli $n^a$ become fields of a
(1+1)-dimensional sigma model on the world sheet. Since
the vector $n^a$ parametrizes the string zero modes,
there is no potential term in this sigma model.  We begin with
the kinetic term \cite{Auzzi}.

To obtain the   kinetic term  we substitute our solution, which depends
on the moduli $ n^a$,
in the action (\ref{model}) assuming  that
the fields acquire a dependence on the coordinates $x_k$ via $n^a(x_k)$.
Then we arrive at the $O(3)$  sigma model\,\footnote{We skip here
some details of derivation.
The interested reader is referred to Refs.~\cite{Auzzi,ShifmanYung}.}
\beq
S^{(1+1)}= \frac{ \beta}{2}\,   \int d t\, dz \,
\left(\pt_k\,  n^a\right)^2\,,\qquad \vec n^2 =1\,,
\label{o3}
\eeq
where the coupling constant $\beta$ is given by a normalizing integral
\beq
\beta=
\frac{2\pi}{g_2^2}\,  \int_0^{\infty}
dr\left\{-\frac{d}{dr}f_3
+\left(\frac{2}{r}\, f_3^2+\frac{d}{dr}f_3\right)\,\frac{\phi_1^2}{\phi_2^2}
\right\}.
\label{beta}
\eeq
Using the first-order equations for the string profile functions (\ref{foe})
one can see that
the integral  here reduces to a total derivative and given
by the flux of the string  determined by $f_3(0)=1$.
This allows us to conclude that the
sigma-model coupling $\beta$ does not depend on the ratio of
the U(1) and SU(2) coupling constants and is given by
\beq
\beta= \frac{2\pi}{g_2^2}\,.
\label{betag}
\eeq
The two-dimensional coupling constant is determined by the
four-dimensional non-Abelian coupling.

In summary,  the effective world-sheet theory describing dynamics
of  the string orientational moduli is the celebrated
$O(3)$ sigma model (which is the same as $CP^1$).  The symmetry
of this model reflects the presence of the  global SU(2)$_{C+F}$
symmetry in the bulk theory.

The relation between the four-dimensional and two-dimensional coupling
constants (\ref{betag}) is obtained  at the classical level. In quantum theory
both couplings run. So we have to specify a scale at which the relation
(\ref{betag}) takes place. The two-dimensional $CP(1)$ model
(\ref{o3}) is
an effective low-energy theory good for the description of
internal string dynamics  at low energies,  much lower than the
inverse thickness of the string which, in turn, is given by $\sqrt{\xi}$. Thus,
$\sqrt{\xi}$ plays the role of a physical ultraviolet (UV) cutoff in
(\ref{o3}).
This is the scale at which Eq.~(\ref{betag}) holds. Below this scale, the
coupling $\beta$ runs according to its two-dimensional renormalization-group
flow.

The sigma model (\ref{o3})
is asymptotically free \cite{Po3}; at large distances (low
energies) it gets into the strong coupling regime.  The  running
coupling constant  as a function of
the energy scale $E$ at one loop is given by
\beq
4\pi \beta = 2\ln {\left(\frac{E}{\Lambda_{CP(1)}}\right)}
+\cdots,
\label{sigmacoup}
\eeq
where $\Lambda_{CP(1)}$ is the dynamical scale of the $CP(1)$
model. As was mentioned above,
the ultraviolet cut-off of the sigma model at hand
is determined by  $\sqrt{\xi}$.
Hence,
\beq
\Lambda^2_{CP(1)} = \xi \, e^{-\frac{8\pi^2}{g^2_2}} .
\label{lambdasig}
\eeq
Note that in the bulk theory, due to the VEV's of
the squark fields, the coupling constant is frozen at
$\sqrt{\xi}$. There are no logarithms in the bulk theory
 below this scale. Below $\sqrt{\xi}$ the logarithms of the
world-sheet  theory take over.

At small values of the deformation parameter
$\mu_2$,
$$\mu_2\ll\sqrt{\xi}\,,$$
the coupling constant $g_2$ of
the four-dimensional bulk theory is determined by the scale $\Lambda$ of
the \ntwo theory. Then Eq.~(\ref{lambdasig}) gives \cite{ShifmanYung}
\beq
\Lambda_{CP(1)}=\Lambda\,,
\label{cpscale1}
\eeq
where we take into account that the first coefficient of the $\beta$ function
equals to 2 both in \ntwo limit of the four-dimensional bulk theory and
in the two-dimensional $CP(1)$ model.

\begin{figure}
\epsfxsize=8cm
\centerline{\epsfbox{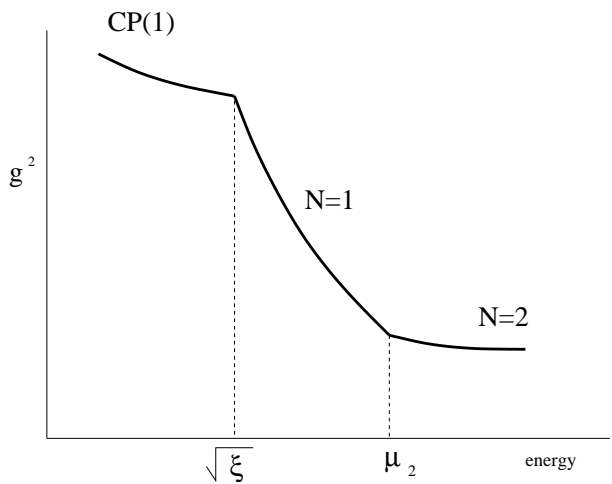}}
\caption{
Renormalization-group (RG) flow in the theory (\ref{model}) at large $\mu_2$.
 Flow in four dimensions is represented
by \none and \ntwo curves, while $CP(1)$ marks the RG flow in
the two-dimensional $CP(1)$ model.}
\label{rgflow}
\end{figure}

Instead, in the limit of large $\mu_2$,
$$\mu_2\gg\sqrt{\xi}\,,$$
the coupling constant $g_2$ of
the  bulk theory is determined by the scale $\Lambda_{{\cal N}=1}$ of
the \none SQCD (\ref{noneqcd}), as shown in Eq.~(\ref{lambdanone}). In this limit
Eq.~(\ref{lambdasig}) gives
\beq
\Lambda_{CP(1)}=\frac{\Lambda_{{\cal N}=1}^2}{\sqrt{\xi}}\,,
\label{cpscale2}
\eeq
where we take into account  that the first coefficient of the $\beta$ function
in \none  SQCD equals to four. The renormalization group flow in our
theory at $\mu_2\gg \sqrt{\xi}$
is schematically presented in Fig.~\ref{rgflow}.

\section{Fermion zero modes}
\label{4}
\renewcommand{\theequation}{\thesection.\arabic{equation}}
\setcounter{equation}{0}

Technically, this is a key section of the
present work. Let us start from the
\ntwo theory (\ref{model}) with the breaking parameters
set to zero, $\mu_i=0$.
Our string solution is 1/2 BPS-saturated. This means that four
supercharges, out of eight of the four-dimensional theory
(\ref{model}), act trivially
on the string solution
(\ref{sna}). The remaining four supercharges generate four fermion zero
modes which we call supertranslational modes because they are
superpartners to  two translational zero
modes. The corresponding four fermionic moduli
are superpartners to the coordinates $x_0$ and
$y_0$ of the string center. The  supertranslational
fermion zero modes were found in Ref.~\cite{VY}.
As a matter of fact, they were found for the U(1) ANO string in
\ntwo  theory\,\footnote{We will review them  and study their
deformation in $\mu$-deformed QED (\ref{qed}) in Sect.~\ref{7}.}
but the transition to the model at hand is absolutely
straightforward. We will not dwell on this procedure
here.

Instead, we will focus below on four {\em additional} fermion zero modes
which arise only for the non-Abelian strings.
They are superpartners of the bosonic orientational moduli $n^a$;
therefore,  we will refer to these modes as superorientational.
In the \ntwo limit these modes were obtained in \cite{ShifmanYung}.
If we switch on supersymmetry (SUSY) breaking parameters $\mu_i$ the number
of supercharges in the four-dimensional bulk theory drops to four.
The 1/2 BPS string would have two superorientational fermion zero modes
in this theory. However, our string is a descendant of \ntwo theory where
it has four superorientational zero modes. Clearly the number of zero
modes cannot jump as we switch on  parameters $\mu_i$,
at least at small $\mu$. This number is determined by index theorems.
Thus, it
is clear that (at least at small $\mu$) our string has a
set of superorientational fermion
zero modes twice bigger than algebra tells.
In this section we elaborate the issue of four  zero modes explicitly
at small and large $\mu$
while  in Sect.~\ref{5} we will
study the impact of their presence  on the $CP(1)$
model on the string world sheet. To begin with, in Sect.~\ref{41} we review
these modes in the \ntwo limit and then examine what happens to them in
the deformed bulk theory.

\subsection{\ntwo limit}
\label{41}

The  fermionic part of the action  of  the model (\ref{model})
is
\beqn
S_{\rm ferm}
&=&
\int d^4 x\left\{
\frac{i}{g_2^2}\bar{\lambda}_f^a \bar{D}\hspace{-0.65em}/\lambda^{af}+
\frac{i}{g_1^2}\bar{\lambda}_f \bar{\pt}\hspace{-0.65em}/\lambda^{f}
+ {\rm Tr}\left[\bar{\psi} i\bar\nabla\hspace{-0.65em}/ \psi\right]
+ {\rm Tr}\left[\tilde{\psi} i\nabla\hspace{-0.65em}/ \bar{\tilde{\psi}}
\right]\right.
\nonumber\\[3mm]
&+&
\frac{i}{\sqrt{2}}\,{\rm Tr}\left[ \bar{q}_f(\lambda^f\psi)+
(\tilde{\psi}\lambda_f)q^f +(\bar{\psi}\bar{\lambda}_f)q^f+
\bar{q}^f(\bar{\lambda}_f\bar{\tilde{\psi}})\right]
\nonumber\\[3mm]
&+&
\frac{i}{\sqrt{2}}\,{\rm Tr}\left[ \bar{q}_f\tau^a(\lambda^{af}\psi)+
(\tilde{\psi}\lambda_f^a)\tau^aq^f +(\bar{\psi}\bar{\lambda}_f^a)\tau^aq^f+
\bar{q}^f\tau^a(\bar{\lambda}^{a}_f\bar{\tilde{\psi}})\right]
\nonumber\\[3mm]
&+&
\frac{i}{\sqrt{2}}\,{\rm Tr}\left[\tilde{\psi}\left(a+a^a\tau^a\right)\psi
\right]
+\frac{i}{\sqrt{2}}\,{\rm Tr}\left[\bar{\psi}\left(a+a^a\tau^a\right)
\bar{\tilde{\psi}}\right]
\nonumber\\[3mm]
&-&
\left. \frac{\mu_1}2 (\lambda^2)^2
-\frac{\mu_2}2 (\lambda^{a2})^2
\right\}\,,
\label{fermact}
\eeqn
where the  matrix color-flavor notation is used for
matter fermions $(\psi^{\alpha})^{kA}$ and $(\tilde{\psi}^{\alpha})_{Ak}$
and the traces are performed
over the color--flavor indices. Contraction of the spinor indices is assumed
inside all parentheses, for instance,
$$(\lambda\psi)\equiv \lambda_{\alpha}\psi^{\alpha}\,.$$
We write the squark fields in (\ref{fermact}) as   doublets of SU(2)$_{R}$
group which is present in \ntwo theory, $q^f=(q,\bar{\tilde{q}})$. Here
$f=1,2$ is the SU(2)$_R$ index
which labels two supersymmetries of the bulk theory in the \ntwo limit.
Moreover,
$\lambda^{\alpha f}$ and $(\lambda^{\alpha f})^a$ stand for
the gauginos of the U(1) and SU(2) groups, respectively.
Note that the last two  terms are \none deformations
in the fermion sector of the theory induced by the breaking parameters $\mu_i$.
They involve only $f=2$ components of $\lambda$'s explicitly breaking
the SU(2)$_{R}$ invariance.

Next, we put $\mu_i=0$ and
apply the general method which was designed in \cite{ShifmanYung}
to generate superorientational fermion zero modes of
the non-Abelian string  in the \ntwo case.
In Ref.~\cite{VY} it was shown that the four supercharges selected
by the conditions
\beq
\epsilon^{11}=0\,, \qquad
\epsilon^{22}=0
\label{trivQ}
\eeq
act trivially on the BPS string in the theory with the
Fayet--Iliopoulos $D$ term.
Here $\epsilon^{\alpha f}$ are parameters of
the SUSY transformation.

Now, to generate the superorientational
fermion zero modes  the following method was used in
\cite{ShifmanYung}. Assume that the
orientational moduli  $n^a$  in the string solution
(\ref{sna}) have a slow dependence on the
world-sheet coordinates $x_0$ and $x_3$ (or $t$ and $z$).
Then the four supercharges selected by
the conditions (\ref{trivQ}) (namely, $\epsilon^{12}$, $\epsilon^{21}$
and their complex conjugates)
no longer  act trivially. Instead,
their action now gives fermion fields proportional to
$x_0$ and $x_3$ derivatives of $n^a$.
This is exactly what one expects from the residual \ntwo supersymmetry
in the world-sheet  theory.
The above four supercharges
generate the world-sheet supersymmetry in the \ntwo
two-dimensional $CP(1)$ model,
\beqn
\delta \chi^a_1
&=&
i\sqrt{2} \left[\left( \pt_0 +i\pt_3\right) n^a \, \varepsilon_2
+\varepsilon^{abc}n^b\left( \pt_0 +i\pt_3\right) n^c\,\eta_2\right]\,,
\nonumber\\[3mm]
\delta \chi^a_2
&=&
i\sqrt{2} \left[\left( \pt_0 -i\pt_3\right)  n^a \, \varepsilon_1
+\varepsilon^{abc}n^b\left( \pt_0 -i\pt_3\right) n^c\,\eta_1\right]\, ,
\label{susy2c}
\eeqn
where $\chi^a_{\alpha}$ ($\alpha=1,2$ is the spinor index) are real
two-dimensional fermions of the $CP(1)$ model. They are
superpartners of $n^a$ and subject to
orthogonality condition $n^a\chi^a_{\alpha}=0$. Real
parameters of \ntwo two-dimensional  SUSY transformation
$\varepsilon_{\alpha}$ and $\eta_{\alpha}$ are identified
with  the parameters of the four-dimensional SUSY
transformations (with the  constraint (\ref{trivQ})) as
\beqn
\varepsilon_1 -i\eta_1
&=&
 \epsilon^{21}\,,
\nonumber\\[3mm]
\varepsilon_2+i\eta_2
&=&
-\epsilon^{12}\, .
\label{epsilon24}
\eeqn
The world-sheet supersymmetry was used to
reexpress the fermion fields obtained upon the action of
these four supercharges
in terms of the (1+1)-dimensional fermions. This procedure  gives
us the superorientational fermion zero modes \cite{ShifmanYung},
\beqn
\bar{\psi}_{Ak\dot{2}}
& = &
\left(\frac{\tau^a}{2}\right)_{Ak}
\frac1{2\phi_2}(\phi_1^2-\phi_2^2)
\left[
\chi_2^a
+i\varepsilon^{abc}\, n^b\, \chi^c_2\,
\right]\, ,
\nonumber\\[3mm]
\bar{\tilde{\psi}}^{kA}_{\dot{1}}
& = &
\left(\frac{\tau^a}{2}\right)^{kA}
\frac1{2\phi_2}(\phi_1^2-\phi_2^2)
\left[
\chi_1^a
-i\varepsilon^{abc}\, n^b\, \chi^c_1\,
\right]\, ,
\nonumber\\[5mm]
\bar{\psi}_{Ak\dot{1}}
& = &
0\, , \qquad
\bar{\tilde{\psi}}^{kA}_{\dot{2}}= 0\, ,
\nonumber\\[4mm]
\lambda^{a22}
& = &
\frac{i}{\sqrt{2}}\frac{x_1+ix_2}{r^2}
f_3\frac{\phi_1}{\phi_2}
\left[
\chi^a_1
-i\varepsilon^{abc}\, n^b\, \chi^c_1
\right]\, ,
\nonumber\\[4mm]
\lambda^{a11}
& = &
\frac{i}{\sqrt{2}}\frac{x_1-ix_2}{r^2}
f_3\frac{\phi_1}{\phi_2}
\left[
\chi^a_2
+i\varepsilon^{abc}\, n^b\, \chi^c_2
 \right]\,,
\nonumber\\[4mm]
\lambda^{a12}
& = & 0
  \, ,\qquad  \lambda^{a21}= 0\,,
\label{zmodes}
\eeqn
where the dependence on $x_i$ is encoded in the string profile
functions, see (\ref{sna}).

Now we will directly verify that the zero modes (\ref{zmodes}) satisfy
the Dirac equations
of motion. From the fermion action of the model (\ref{fermact}) we get
the relevant Dirac equations for $\lambda^a$,
\beq
\frac{i}{g_2^2} \bar{D}\hspace{-0.65em}/\lambda^{af}
+\frac{i}{\sqrt{2}}\,{\rm Tr}\left(
\bar{\psi}\tau^a q^f+
\bar{q}^f\tau^a\bar{\tilde{\psi}}\right)-
\mu_2\delta^f_2\bar{\lambda}_2^a=0\,,
\label{dirac1}
\eeq
while  for the matter fermions
$$
i\nabla\hspace{-0.65em}/ \bar{\psi}+\frac{i}{\sqrt{2}}\left[\bar{q}_f\lambda^f
-(\tau^a\bar{q}_f)\lambda^{af}+
(a-a^a\tau^a)\tilde{\psi}\right]=0\, ,
$$
\beq
i\nabla\hspace{-0.65em}/ \bar{\tilde{\psi}}+
\frac{i}{\sqrt{2}}\left[\lambda_f q^f
+\lambda^{a}_f(\tau^aq^f)+
(a+a^a\tau^a)\psi\right]=0\, .
\label{dirac2}
\eeq
Next,  we
substitute the orientational fermion zero modes (\ref{zmodes}) into these
equations and take the limit $\mu_2=0$. After some algebra one can check
that (\ref{zmodes}) do  satisfy the Dirac equations (\ref{dirac1}) and
(\ref{dirac2})
provided the first-order equations for string profile functions (\ref{foe})
are fulfilled.

It is instructive to check that the zero
modes (\ref{zmodes}) do produce
the fermion part of the   \ntwo two-dimensional $CP(1)$ model.
To this end we return to the usual assumption that
the fermion collective coordinates  $\chi^a_{\alpha}$ in
Eq.~(\ref{zmodes}) have an adiabatic  dependence on the world-sheet
coordinates $x_k$ ($k=0,3$). This is quite similar to the procedure of
Sect.~\ref{32}.
Substituting Eq.~(\ref{zmodes}) in  the fermion kinetic terms
in the bulk  theory (\ref{fermact}),
and taking into account the derivatives of  $\chi^a_{\alpha}$ with
respect to the world-sheet coordinates  we arrive at
\beq
\beta \int d t d z \left\{\frac12 \, \chi^a_1 \, (\pt_0-i\pt_3)\, \chi^a_1
+\frac12 \, \chi^a_2 \, (\pt_0+i\pt_3)\, \chi^a_2
\right\},
\label{fkin}
\eeq
where $\beta$ is given by the same integral (\ref{beta}) as for
the bosonic kinetic term, see Eq.~(\ref{o3}).

We can use the world-sheet \ntwo supersymmetry to reconstruct the
four-fermion interactions inherent to $CP(1)$. The SUSY transformations
in the $CP(1)$ model
have the form (see \cite{NSVZsigma} for a review)
\beqn
\delta \chi^a_1
&=&
i\sqrt{2} \left( \pt_0 +i\pt_3\right) n^a \, \varepsilon_2
+\sqrt{2}\varepsilon_1\,n^a(\chi^a_1\chi^a_2)\,,
\nonumber\\[3mm]
\delta \chi^a_2
&=&
i\sqrt{2} \left( \pt_0 -i\pt_3\right)  n^a \, \varepsilon_1
-\sqrt{2}\varepsilon_2\,n^a(\chi^a_1\chi^a_2)\, ,
\nonumber\\[3mm]
\delta n^a
&=&
\sqrt{2}(\varepsilon_1\chi^a_2+\varepsilon_2\chi^a_1)\,,
\label{susy2d}
\eeqn
where  for simplicity we put $\eta_{\alpha}=0$. Imposing this supersymmetry
leads to the following effective theory on the string world sheet:
\beqn
S_{CP(1)}
&=&
\beta \int d t d z \left\{\frac12 (\pt_k n^a)^2+
\frac12 \, \chi^a_1 \, i(\pt_0-i\pt_3)\, \chi^a_1
\right.
\nonumber\\[3mm]
&+& \left.
\frac12 \, \chi^a_2 \, i(\pt_0+i\pt_3)\, \chi^a_2
-\frac12 (\chi^a_1\chi^a_2)^2
\right\},
\label{ntwocp}
\eeqn
This is indeed the action of the \ntwo $CP(1)$ sigma model.

\subsection{Breaking \ntwo supersymmetry}
\label{42}

Now let us switch on our breaking parameters $\mu_i$. As was discussed
in Sect.~\ref{3},
the bosonic solution for
the non-Abelian string does not change at all. It is still given
by Eq.~(\ref{sna}). However, the fermion zero modes do change.
Now only four supercharges
survive  in the four-dimensional bulk theory. They are  associated with the parameters
$\epsilon^{\alpha 1}$ for $f=1$.
Nevertheless, we still can use the method of
Ref.~\cite{ShifmanYung} reviewed in   Sect.~\ref{41} to generate
superorientational fermion zero modes. Condition (\ref{trivQ}) tells
us that we now have only one complex parameter $\epsilon^{21}$
of SUSY transformations unbroken by the string. This leads to the
presence of two supercharges associated with two real parameters
$\varepsilon_{1}$ and $\eta_{1}$, according to identification (\ref{epsilon24}),
in the world-sheet theory.
Following the same steps which led us to (\ref{zmodes})
and taking into account that the bosonic string solution (\ref{sna})
does  not depend on $\mu_i$  we then obtain
\beqn
\bar{\psi}_{Ak\dot{2}}
& = &
\left(\frac{\tau^a}{2}\right)_{Ak}\,\,
\frac1{2\phi_2}(\phi_1^2-\phi_2^2)
\left[
\chi_2^a
+i\varepsilon^{abc}\, n^b\, \chi^c_2\,
\right]\, ,
\nonumber\\[3mm]
\bar{\psi}_{Ak\dot{1}}
& = &
0\, ,
\nonumber\\[4mm]
\lambda^{a11}
& = &
\frac{i}{\sqrt{2}}\frac{x_1-ix_2}{r^2}
\, f_3\, \frac{\phi_1}{\phi_2}
\left[
\chi^a_2
+i\varepsilon^{abc}\, n^b\, \chi^c_2
\right]\,,
\nonumber\\[4mm]
\lambda^{a21}
& = & 0
\, .
\label{nonemodes}
\eeqn

We see that reduced supersymmetry generates for  us  only two fermion
superorientational modes parametrized by
the  two-dimensional fermion field
$\chi_2^a$.
This was expected, of course.
The modes proportional to $\chi^a_1$ do not appear. This is because
$\chi^a_1$ is related to the
SUSY transformations generated by $\epsilon^{12}$
(see (\ref{susy2c}) and (\ref{epsilon24})) which is no longer present in the
deformed bulk theory. One can easily check that zero modes
(\ref{nonemodes}) still satisfy the Dirac equations of motion (\ref{dirac1}),
(\ref{dirac2})
just because the parameter $\mu_2$ does not enter the equations for
$\lambda^{\alpha 1}$ and $\bar{\psi}$.

It is clear, however, that the other two fermion zero modes proportional to
$\chi_1$ do not disappear. They are just modified and can no longer
be obtained by supersymmetry. To find them we have to
actually solve the Dirac
equations (\ref{dirac1}), (\ref{dirac2}). In this  section
we consider small $\mu_2$
and develop perturbation theory for (\ref{dirac1}), (\ref{dirac2}).
In  Sect.~\ref{43} we treat the large $\mu_2$ limit.

We can solve (\ref{dirac1}), (\ref{dirac2})
order by order in $\mu_2$. Say, if we take
(\ref{zmodes}) for the zeroth- order approximation and substitute $\lambda^{22}$
from (\ref{zmodes}) into the last term in Eq.~(\ref{dirac1}) we generate
fermion zero modes to the first order in $\mu_2$. Let us actually do this.

First we note that
\beq
\bar{\tilde{\psi}}^{kA}_{\dot{2}}=0, \;\;\;\; \lambda^{a12}=0\, .
\label{dot212}
\eeq
They vanish in the zeroth order (see (\ref{zmodes})) and,
as follows from Eqs.~(\ref{dirac1}) and (\ref{dirac2}), are not generated in
any order in $\mu_2$. It is also easy to check that the remaining
fermion fields  have the following form:
\beqn
\lambda^{a22}
&=&
\frac{x_1+ix_2}{r}\,\lambda_{+}(r)\,\left[\chi_1^a-
i\varepsilon^{abc}n^b\chi_1^c\right] + \lambda_{-}(r)\,\left[\chi_1^a+
i\varepsilon^{abc}n^b\chi_1^c\right]\, ,
\nonumber\\[4mm]
\bar{\tilde{\psi}}^{kA}_{\dot{1}}
&=&
\psi_{+}(r)\,
\left(\frac{\tau^a}{2}\right)^{kA}\,
\left[\chi_1^a-
i\varepsilon^{abc}n^b\chi_1^c\right]
\nonumber\\[4mm]
&+&
\frac{x_1-ix_2}{r}\,\psi_{-}(r)\,
\left(\frac{\tau^a}{2}\right)^{kA}\,
\left[\chi_1^a+
i\varepsilon^{abc}n^b\chi_1^c\right].
\label{fprofile}
\eeqn
Here we introduced four profile functions $\lambda_{\pm}$ and $\psi_{\pm}$
parametrizing the fermion fields $\lambda^{22}$ and $\bar{\tilde{\psi}}_{\dot{1}}$.
The functions $\lambda_{+}$ and $\psi_{+}$ are expandable
in even powers of
$\mu_2$ while the functions  $\lambda_{-}$ and $\psi_{-}$    in
odd powers of $\mu_2$.

Substituting (\ref{fprofile}) into the Dirac equations (\ref{dirac1}),
(\ref{dirac2}) we get following equations for fermion profile functions:
\beqn
&&\frac{d}{dr}\psi_{+} -\frac1{2r}(f-f_3)\psi_{+}+i\sqrt{2}\,\phi_1\,
\lambda_{+}=0\, ,
\nonumber\\[3mm]
&-&\frac{d}{dr}\lambda_{+}-\frac1r\lambda_{+}+\frac{f_3}{r}\lambda_{+}
+i\frac{g^2_2}{\sqrt{2}}\,\phi_1\,\psi_{+} +g^2_2\mu_2\,\lambda_{-}=0\,,
\nonumber\\[3mm]
&&\frac{d}{dr}\psi_{-} +\frac1r\psi_{-}
-\frac1{2r}(f+f_3)\psi_{-}+i\sqrt{2}\,\phi_2\,\lambda_{-}=0,
\nonumber\\[3mm]
&-&\frac{d}{dr}\lambda_{-}-\frac{f_3}{r}\lambda_{-}
+i\frac{g^2_2}{\sqrt{2}}\,\phi_2\,\psi_{-} +g^2_2\mu_2\,\lambda_{+}=0\,.
\label{fermeqs}
\eeqn
The  leading contributions to the $\mu$ even solutions to these equations is
\beq
\lambda_{+}=\frac{i}{\sqrt{2}}\frac{f_3}{r}\frac{\phi_1}{\phi_{2}}
+O(\mu_2^2),\,\,\,\,\,
\psi_{+}=\frac{1}{2\phi_2}\left(\phi_1^2-\phi_2^2\right)+O(\mu_2^2)\,,
\label{zeroorder}
\eeq
where we express the zeroth-order fermion modes $\lambda^{22}$ and
$\bar{\tilde{\psi}}_{\dot{1}}$ (\ref{zmodes}) in terms of
the fermion profile functions.
Substituting (\ref{zeroorder}) into the last equation in (\ref{fermeqs})
we can solve for the leading contributions to the $\mu$ odd profile functions.
They can be expressed in terms of the string profile functions as follows:
\beqn
\lambda_{-}
&=&
\mu_2\,\frac{i}{2\sqrt{2}}\,\left[(f_3-1)\frac{\phi_2}{\phi_1}
+\frac{\phi_1}{\phi_2}\right]+O(\mu_2^3)\,,
\nonumber\\[3mm]
\psi_{-}
&=&
\mu_2\frac{r}{4\phi_1}\left(\phi_1^2-\phi_2^2\right)+
O(\mu_2^3)\, .
\label{firstorder}
\eeqn
Using the boundary conditions (\ref{fbc}) and (\ref{phibc})
for the string profile functions  it is easy to check that these solutions
vanish at $r\to\infty$ and are non-singular at $r=0$.

We conclude that the number of
the superorientational zero modes of the
non-Abelian string does not jump as we switch on
the deformation parameters
$\mu_i$. We keep all  four zero modes parametrized by $\chi_1^a$ and
$\chi_2^a$. The modes proportional to $\chi_1^a$ are now modified.
Still, we can find them order by order in $\mu_2$ by solving the Dirac
equations (\ref{fermeqs}). As was mentioned in  Sect.~\ref{intro}
(and will
be explained in detail in   Sect.~\ref{5}) the four fermion zero modes imply
\ntwo supersymmetry in the two-dimensional world-sheet sigma model
(four supercharges).
On the other hand,
\ntwo  supersymmetry in the bulk theory is broken down to
\none ( four supercharges). Thus, we do observe
enhancement of supersymmetry on the
string world sheet.

On general grounds one might
expect a breaking of the enhanced  world-sheet
supersymmetry at some critical value $\mu_i^*$. What could happen is
the fermion
zero modes associated with $\chi_1^a$
could become non-normalizable at some
value of $\mu_2$. Clearly,
one would not be able to see  the
loss of normalizability in  perturbation theory
in $\mu_2$. In Sect.~\ref{43} we will examine the limit of large $\mu_i$
and show that the fermion modes (\ref{fprofile}) become non-normalizable
only at $\mu_2\to\infty$.

\subsection{The large $\mu$ limit}
\label{43}

Let us dwell on the limit of large $\mu_2$, or,
more explicitly\,\footnote{The parameter $\omega_1$ does not enter Eqs.~(\ref{fermeqs}),
therefore we can ignore it.}
\beq
\omega_2\gg 1\,,
\label{largemulim}
\eeq
see (\ref{omega}). As was explained in Sect.~\ref{bth} the
fields  $a^a$ (as well as their fermion counterparts $\lambda^{a\alpha 2}$)
become heavy and can be integrated out. The low-energy theory for
the SU(2) sector contains one massive SU(2) gauge multiplet, with mass
\beq
m_0\equiv m_{{\rm SU}(2)}=g_2\sqrt{\xi}\,,
\label{gmass}
\eeq
see (\ref{wmass}), and three ($a=1,2,3$) chiral light multiplets, with mass
\beq
m_L\equiv m^{-}_{{\rm SU}(2)}=\frac{\xi}{\mu_2}\, .
\label{lmass}
\eeq

Integrating out heavy fields can be carried out in  superpotentials
(\ref{superpot}), (\ref{superpotbr}), as in \cite{KSS,GVY,VY}, or directly
in the component Lagrangian. One just drops
the kinetic terms for the heavy fields
and solves algebraic equations for these fields. We do it in the
fermion  sector of the theory in the Dirac equations  (\ref{dirac1})
for $\lambda^{a\alpha 2}$. More exactly, we  get  expressions for
the $\lambda$-profile functions in terms of the $\psi$-profile functions
from the first and the third equations in (\ref{fermeqs}). Namely,
\beqn
\lambda_{+}
&=&
\frac{i}{\sqrt{2}\phi_1}\left[\frac{d}{dr}\psi_{+} -
\frac1{2r}(f-f_3)\psi_{+}\right],
\nonumber\\[3mm]
\lambda_{-}
&=&
\frac{i}{\sqrt{2}\phi_2}\left[
\frac{d}{dr}\psi_{-} +\frac1r\psi_{-}-\frac1{2r}(f+f_3)\psi_{-}\right].
\label{lambdapsi}
\eeqn
Dropping the kinetic term for
$\lambda$'s  in the second and the fourth equations in  (\ref{fermeqs}) and
substituting (\ref{lambdapsi}) in these equations we arrive at
\beqn
&&\frac{d}{dr}\psi_{+} -\frac1{2r}(f-f_3)\psi_{+}+m_L\,\frac{\phi_1\phi_2}{\xi}
\,\psi_{-}=0\, ,
\nonumber\\[3mm]
&&\frac{d}{dr}\psi_{-} +\frac1r\psi_{-}-\frac1{2r}(f+f_3)\psi_{-}+
m_L\,\frac{\phi_1\phi_2}{\xi}\,\psi_{+}=0\,,
\label{psieqs}
\eeqn
where $m_L$ is the light mass given in Eq.~(\ref{lmass}).

Now observe that long-range tails of the solutions to these equations are
determined by the small mass $m_L$, while the string profile functions
$f$ and $f_3$ are important at much smaller distances $R\sim 1/m_0$.
This key observation allows us to solve Eqs.~(\ref{psieqs})
analytically. We will treat separately two domains: (i) large $r$,
$$r\gg 1/m_0$$ and (ii)
intermediate $r$,
$$r\le 1/m_0\,.$$

\vspace{3mm}

{\bf Large-\boldmath{$r$} domain, \boldmath{$r\gg 1/m_0$}}

\vspace{2mm}
\noindent
In this domain
we can drop the terms in (\ref{psieqs}) containing $f$ and $f_3$ and use the
first equation
to express $\psi_{-}$ in terms of $\psi_{+}$. We then get
\beq
\psi_{-}=- \frac1{m_L}\frac{d}{dr}\psi_{+}\, .
\label{psipm}
\eeq
Substituting this into the second equation in (\ref{psieqs}) we obtain
\beq
\frac{d^2}{dr^2}\psi_{+}+\frac1r\frac{d}{dr}\psi_{+}-m_L^2\psi_{+}=
0\,.
\label{psipluseq}
\eeq
This is a well-known equation for a free field with mass $m_L$
in the radial coordinates. Its solution is  well-known
too\,\footnote{Equation~(\ref{psipluseq}) determines
the profile function $\psi_{+}$ up to an overall normalization constant.
This constant is included in the normalization of
the two-dimensional fermion
field $\chi_1^a$. We will discuss this normalization in Sect.~\ref{5}.}
\beq
\psi_{+}=m_L\, K_0(m_L r),
\label{psiplus}
\eeq
where $K_0 (x)$ is the imaginary argument Bessel function. At infinity
it  falls-off  exponentially,
\beq
K_0(x)\sim \frac{e^{-x}}{\sqrt{x}}\,,
\eeq
while at $x\to 0$ it has the
logarithmic behavior,
\beq
K_0(x)\sim \ln{\frac1x}\, .
\label{log}
\eeq
Taking into account (\ref{psipm}) we get the solutions for
the  fermion profile
functions at $r\gg 1/m_0$,
\beq
\psi_{+}=m_L\, K_0(m_L r)\,,\qquad \psi_{-}=- \frac{d}{dr}K_0(m_L r)\, .
\label{psi}
\eeq
In particular, at $r\ll 1/m_L$ we have
\beq
\psi_{+}\sim m_L\, \ln\, {\frac1{m_L r}}\,,
\qquad
\psi_{-}\sim\frac1r  \,.
\label{psizero}
\eeq

\vspace{3mm}

{\bf Intermediate-\boldmath{$r$} domain, \boldmath{$r\le 1/m_0$}}

\vspace{2mm}
\noindent
In this  domain we neglect
small mass terms in (\ref{psieqs}). We then arrive at
\beqn
&&\frac{d}{dr}\psi_{+} -\frac1{2r}(f-f_3)\psi_{+} =0\,,
\nonumber\\[3mm]
&&\frac{d}{dr}\psi_{-} +\frac1r\psi_{-}-\frac1{2r}(f+f_3)\psi_{-}=0\,.
\label{smallreqs}
\eeqn
These equations are identical to those for the string profile
functions, see (\ref{foe}). Therefore, their solutions are known,
\beq
\psi_{+}=c_1\phi_2\,,
\qquad
\psi_{-}=\frac{c_2}{r}\phi_1\,,
\label{spsi}
\eeq
up to normalization constants $c_{1,2}$.
To fix these constants  we match the long-distance behavior
in (\ref{spsi}) with the short-distance behavior of the solutions
in the domain $r\gg 1/m_0$ given in (\ref{psizero}).
This gives the fermion profile functions at intermediate $r$,
\beq
\psi_{+}=\frac{m_L\ln\, {(m_0/m_L)}}{\sqrt{\xi}}\,\phi_2\,,
\qquad
\psi_{-}=\frac{1}{r\sqrt{\xi}}\,\phi_1\, .
\label{sdpsi}
\eeq

Equations~(\ref{psi}) and (\ref{sdpsi}) present our final result for
the fermion profile
functions in the limit of large $\mu_2$. They determine two fermion
superorientational
zero modes proportional to $\chi_1^a$ via Eq.~(\ref{fprofile}). The main
feature of these modes is the presence of the long-range tails determined
by the {\em small} mass $m_L$. Neither bosonic string solution
(\ref{sna}) nor two other  superorientational fermion zero modes
(\ref{nonemodes}) determined by \none supersymmetry
have these logarithmic long-range tails\,\footnote{Here the word {\em logarithmic}
is used in a somewhat Pickwick sense. More precisely
one should say that the large-distance behavior of the long-range tails
is such that the corresponding normalization factors diverge logarithmically.
This divergence is cut off at $m_L^{-1}.$}.

\section{Effective world-sheet theory in the large-\boldmath{$\mu$} limit}
\label{5}
\renewcommand{\theequation}{\thesection.\arabic{equation}}
\setcounter{equation}{0}

To fully specify the fermion  sector of the world-sheet sigma model
we substitute the fermion zero modes (\ref{psi}), (\ref{sdpsi}) and
(\ref{nonemodes}) into the fermion action (\ref{fermact}),
much in the same way we
did  in Sect.~\ref{41} in the \ntwo limit. Then instead of
Eq.~(\ref{ntwocp}) we get
\beqn
S_{1+1}
&=&
\beta \int d t d z \left\{\frac12 (\pt_k n^a)^2+
I_f\frac12 \, \chi^a_1 \, i(\pt_0-i\pt_3)\, \chi^a_1
\right.
\nonumber\\[3mm]
&+& \left.
\frac12 \, \chi^a_2 \, i(\pt_0+i\pt_3)\, \chi^a_2
- I_f\frac12 (\chi^a_1\chi^a_2)^2
\right\}.
\label{cp}
\eeqn
Here $I_f$ is the normalization integral for
the  deformed fermion zero modes
(\ref{psi}) and (\ref{sdpsi}). Its leading behavior at large $\mu_2$ is
given by
\beq
I_f = 2g^2_2\int rdr\left(|\psi_{+}|^2 +|\psi_{-}|^2\right)\sim g_2^2\,
\ln\,{\left(\frac{m_0}{m_L}\right)}\,
\eeq
coming from $\psi_{-}$.
Substituting the mass values from (\ref{gmass}) and (\ref{lmass})
we then obtain
\beq
I_f\sim g_2^2\,\ln\,{\left(\frac{g_2\mu_2}{\sqrt{\xi}}\right)}\,.
\label{if}
\eeq
Note that the calculation actually gives us only the bilinear fermion terms
in (\ref{cp}). We fix the coefficient in front of the quartic term
using \none supersymmetry on the world sheet generated by parameters
$\varepsilon_1$ and $\eta_1$, see (\ref{susy2d}). This supersymmetry is
necessarily present in our world-sheet theory. In particular, it relates the
coefficient in front of the kinetic term for $\chi^a_1$ and the one
in front of the quartic term.

Next, we absorb the normalization integral $I_f$ in the definition
of the fermion fields $\chi_1^a$. As a result, we arrive at
the $CP(1)$ model
(\ref{ntwocp}). This model has \ntwo supersymmetry in two dimensions
(four supercharges). We thus confirm enhancement of supersymmetry in
our effective  theory on the string world sheet. As was explained in
Sect.~\ref{intro}, this result could be expected on general grounds. The
target space of the $CP(1)$ model ($S_2$ sphere) is the K\"ahler manifold. Supersymmetry
on the K\"ahler manifolds requires four supercharges. If our string
is BPS  and the world-sheet theory is local,
the world-sheet supersymmetry {\em must} be enhanced.
The same reasoning was recently used  \cite{Adam2} to
prove enhanced supersymmetry on the world volume of domain walls.

If we started directly from \none SQCD (\ref{noneqcd}) we would
have never
obtained enhanced supersymmetry on the world sheet of
the non-Abelian string.
We would find only two  fermion zero modes (\ref{nonemodes}), while
the other two are non-normalizable. The reason for this is the presence
of the Higgs
branch in (\ref{noneqcd}). Embedding (\ref{noneqcd}) in the deformed
\ntwo theory (\ref{model}) lifts the Higgs branch and makes the
second pair of the fermion zero modes normalizable at any finite $\mu_2$.
This infrared (IR) regularization allows us to obtain \ntwo supersymmetric $CP(1)$
model (\ref{ntwocp}) as an effective theory on the world sheet of
the non-Abelian string.

\section{Limits of applicability}
\label{6-1}
\renewcommand{\theequation}{\thesection.\arabic{equation}}
\setcounter{equation}{0}

\begin{figure}
\epsfxsize=8cm
\centerline{\epsfbox{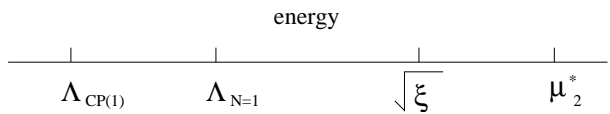}}
\caption{
A spectrum of relevant scales in the limit  $\mu_2\gg \sqrt{\xi}$.}
\label{scales}
\end{figure}

Although the two-derivative term we derived above
is \ntwo supersymmetric for any finite $\mu$ one
should expect the enhanced \ntwo supersymmetry to be broken
at some (large) value of $\mu_2$ due to induced terms
with four or more derivatives. Let us determine this critical value.
To this end let us note that  higher derivative
corrections run in powers of
\beq
\Delta\, \pt_k,
\label{hder}
\eeq
where $\Delta$ is a string transverse size. At small $\mu_2$,
$$\Delta\sim1/\sqrt{\xi}\,.$$
The typical energy scale on the string world
sheet is given by the scale $\Lambda_{CP(1)}$ of the $CP(1)$ model which
is given by (\ref{cpscale1}) at small $\mu_2$. Thus, $$\pt \to\Lambda$$
and
higher derivative corrections in fact run  in powers of $\Lambda/\sqrt{\xi}$.
At small $\mu_2$   higher derivative  corrections are suppressed as
$ \Lambda/\sqrt{\xi}\ll1$, and we can ignore them.
However, as we increase $\mu_2$ the fermion zero modes
(\ref{psi}), (\ref{sdpsi}) acquire long-range tails. This means that an effective
``fermion" thickness of the string grows and becomes
\beq
\Delta\sim \frac{1}{m_L}=\frac{\mu_2}{\xi}.
\label{delta}
\eeq
Higher derivative terms are small if $\Delta\,\Lambda_{CP(1)}\ll 1$.
Substituting
here the scale of the $CP(1)$ model given by (\ref{cpscale2}) at large $\mu_2$
and the scale of \none SQCD (\ref{lambdanone}) we arrive at
\beq
\mu_2\ll\mu_2^{*}\,,
\label{cond}
\eeq
where the critical value of   $\mu_2$ is given by
\beq
\mu_2^{*} = \frac{\xi}{\Lambda_{CP(1)}}=\frac{\xi^{3/2}}
{\Lambda^2_{{\cal N}=1}}\,.
\label{mucrit}
\eeq
If the  condition (\ref{cond}) is met,
the \ntwo CP(1) model gives  a good description of the world-sheet physics.
A spectrum of relevant scales in our theory is shown in Fig.~\ref{scales}.

If we increase $\mu_2$ above the critical value (\ref{mucrit})
the non-Abelian strings
become effectively thick and their  world-sheet dynamics is
no  longer described   by \ntwo $CP(1)$ sigma model. The
higher derivative corrections on the world sheet explode. Since the higher
derivative sector does not respect the enhanced \ntwo supersymmetry
the latter  gets broken down to \none (two supercharges).

Note that the physical reason for the growth of the string thickness $\Delta$
is the presence of the Higgs branch in \none SQCD (\ref{noneqcd}). Although
the classical string solution (\ref{sna}) stays compact, the
presence of the Higgs branch shows up at the quantum level. In particular,
the fermion zero modes  feel  its presence and acquire long-range
logarithmic tails.

Summarizing, the  \ntwo $CP(1)$ model with enhanced supersymmetry is a valid description
of the world-sheet physics  of the non-Abelian string
if the condition (\ref{cond}) is met. Otherwise the \ntwo world-sheet
supersymmetry is broken down to \none by
higher derivative terms. Simultaneously, the string at hand becomes
``thick." By thick we mean that its transverse dimension is
determined by the large parameter $\mu_2/\xi \to \infty$ rather than
by $\xi^{-1/2}$.

\section{Non-Abelian monopoles in \none }
\label{naminnone}
\renewcommand{\theequation}{\thesection.\arabic{equation}}
\setcounter{equation}{0}

Since the \ntwo $CP(1)$ model is the effective
low-energy theory describing
the world-sheet physics  of the non-Abelian string
all consequences of this model ensue,
in particular, two degenerate vacua and a kink
which interpolates between them --- the same kink
that we had in \ntwo  \cite{ShifmanYung} and interpreted as a
(confined) non-Abelian monopole,
the descendent of the 't Hooft--Polyakov monopole \cite{thopo}.

Let us briefly review  the reason for this interpretation
\cite{Tong,ShifmanYung}. We first set to zero
the \ntwo breaking parameters $\mu_i$
in (\ref{model})  and introduce a mass difference $\Delta m$ for
two quark supermultiplets, see \cite{ShifmanYung} for details.
Let us start from the vanishing
FI parameter $\xi$ (i.e. start from the Coulomb
branch). At  $\Delta m \neq 0$ the gauge group SU(2) is broken down
to U(1) by a VEV of the SU(2) adjoint scalar $\langle a^3\rangle  \sim \Delta m$. Thus,
there are 't Hooft-Polyakov monopoles of broken gauge SU(2).
Classically, on the Coulomb branch
their mass is proportional to $| \Delta m \,| /g_2^2$.
 In  the   limit $\Delta m\to 0$ they
become massless, formally, in the classical
approximation. Simultaneously their size become infinite \cite{We}.
The mass and size are stabilized by confinement effects
which are highly quantum. The confinement of monopoles occurs
in  the Higgs phase, at $\xi\neq 0$.

\begin{figure}[h]
\epsfxsize=12cm
\centerline{\epsfbox{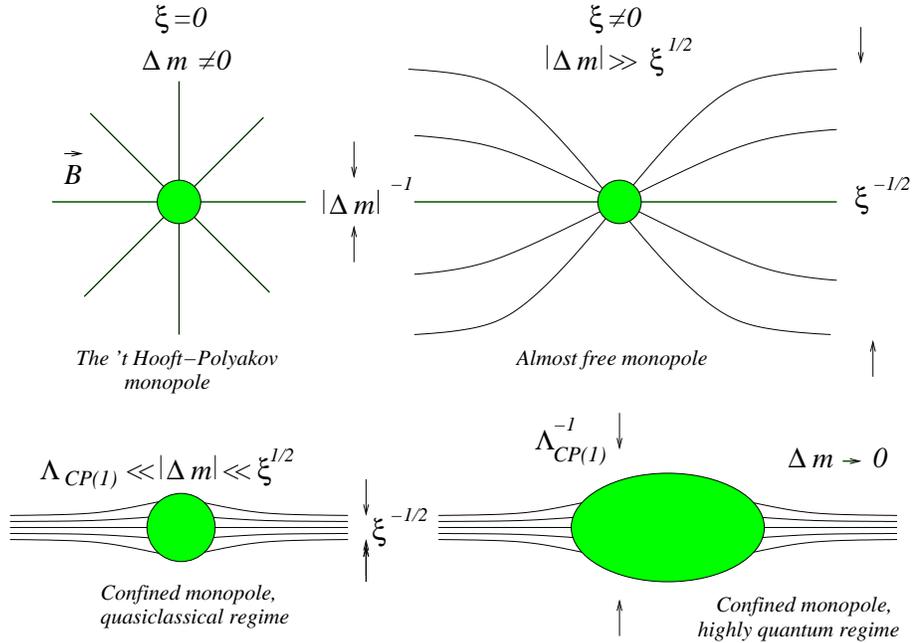}}
\caption{
Various regimes for the monopoles and
flux tubes. The latter case
corresponds to the vanishing $\Delta m$.}
\label{twoabcd}
\end{figure}

A qualitative evolution of the monopoles under consideration
as a function of the  relevant parameters is presented in
Fig.~\ref{twoabcd}.

We begin with the limit $\xi \to 0$ while $\Delta m $ is kept fixed.
Then the corresponding  microscopic theory supports
the conventional (unconfined)
't~Hooft-Polyakov monopoles \cite{thopo}
due to the spontaneous breaking of the {\em gauge} SU(2)
down to U(1),
(the  upper left corner of Fig.~\ref{twoabcd}).

If we allow $\xi$ be non-vanishing but
\beq
|\Delta m | \gg\sqrt{\xi}
\eeq
then the effect which comes into play first is the above
spontaneous breaking of the   gauge SU(2).
Further gauge symmetry breaking, due to $\xi\neq 0$,
which leads to complete Higgsing of the model and the string
formation (confinement of monopoles) is much weaker.
Thus,  we deal here with  the formation
of ``almost"  't~Hooft-Polyakov monopoles, with a typical size
$\sim \left| \Delta m\right| ^{-1}\,.$ Only at much larger distances,
$\sim \xi ^{-1/2}$, the charge condensation enters the game,
and forces the magnetic flux, rather than spreading evenly a l\'a
Coulomb, to form flux tubes (the  upper right corner of
Fig.~\ref{twoabcd}).  There will be two such flux tubes, with the distinct
orientation of the color-magnetic flux ($Z_2$ strings discussed in
Sect. 3.1).
 The monopoles, albeit confined, are weakly confined.

Now, if we further reduce $\left| \Delta m\right| $,
\beq
\Lambda_{CP(1)} \ll \left| \Delta m\right|   \ll \sqrt{\xi}\, ,
\label{ququr}
\eeq
the size of the monopole ($\sim \left| \Delta m\right|^{-1} $) becomes
larger than the transverse size of the attached strings.
The monopole gets squeezed  in earnest by
the strings --- it becomes  a {\em bona fide} confined
monopole (the  lower left corner of  Fig.~\ref{twoabcd}).
A macroscopic description of such monopoles is provided
by  the twisted-mass $CP(1)$ model
on the string world sheet \cite{Tong, ShifmanYung}.
Namely two $Z_2$ strings are interpreted as two vacua of
the $CP(1)$
model while the monopole (string junction of two $Z_2$ strings) is
interpreted as a kink interpolating between these two vacua.

The value of the twisted mass equals $ \Delta m$ while the
size of
the twisted-mass sigma-model kink/confined monopole
is of order of  $ \left| \Delta m\right|^{-1} $.
As we further diminish $\left| \Delta m\right|$
approaching $\Lambda_{CP(1)}$ and then getting  below  $\Lambda_{CP(1)}$,
the size of the monopole grows, and, classically, it would explode.
This is where quantum effects in the world-sheet theory take over.
It is natural to refer to this domain of parameters as the ``regime of
highly quantum dynamics."
While the thickness of the string (in the transverse direction) is
$\sim \xi ^{-1/2}$, the
$z$-direction size of the kink  representing the confined
monopole in the highly quantum regime is much larger, $\sim
\Lambda_{CP(1)}^{-1}$, see the  lower right corner of
Fig.~\ref{twoabcd}.

In \cite{ShifmanYung} the first order equations for 1/4 BPS string
junction of two $Z_2$ strings were explicitly solved and the solution
shown to correspond to a kink solution of
the two-dimensional $CP(1)$
model. Moreover, it was shown that the mass of the monopole matches
the mass of the $CP(1)$-model kink  both in the quasiclassical ($\Delta m\gg
\Lambda_{CP(1)}$)   and quantum  ($\Delta m \ll \Lambda_{CP(1)}$)
limits.

Thus, at zero $\Delta m$ we still have a confined ``monopole" stabilized
by quantum effects in the world-sheet $CP(1)$ model (interpreted
as a kink). Now we can switch on the \ntwo breaking parameters $\mu_i$.
If we keep $\mu_2$ less than the critical value (\ref{mucrit})
the effective world-sheet description of the non-Abelian string is
still given by the \ntwo $CP(1)$ model. This model obviously
still has two vacua
which should be interpreted as two elementary non-Abelian
strings in the quantum regime, and a BPS kink
can interpolate between these vacua. This kink
should still be interpreted as a non-Abelian
confined monopole/string junction.
Its mass and inverse size is determined by $\Lambda_{CP(1)}$ which
in the limit of large $\mu_2$ is given by Eq.~(\ref{cpscale2}).

This kink--monopole is half-critical considered
from the standpoint of the $CP(1)$ model
(i.e. two supercharges conserved).
Thus, we observe
supersymmetry   enhancement
at the next level too. In fact, this is ``supersymmetry emergence"
rather than enhancement, since in the bulk \none theory there is
no such thing as the monopole central charge!
Indeed, in the \ntwo model \cite{SW}
there exists a ``monopole" central charge \cite{HLS}
which implies, in turn, the critical nature of
the 't Hooft--Polyakov monopole. By appropriately varying  parameters of the model
one can trace continuous evolution of the conventional (unconfined)
$\mbox{'t~Hooft}$--Polyakov monopole into a
weakly confined monopole and then into 1/2-BPS non-Abelian confined kink--monopole
in a highly quantum regime.

In the \none model at hand the monopole central charge cannot
exist for symmetry reasons, and one cannot expect BPS-saturated
$\mbox{'t~Hooft}$--Polyakov monopoles.
On the other hand, the kink central charge
certainly exists in the two-dimensional
superalgebra \cite{LoShi} pertinent to the $CP(1)$ model.
Here we encounter
the notion of a central charge that exist in the low-energy moduli
theory but cannot be lifted to the bulk theory as a matter of
principle. A similar phenomenon does
actually occur in the domain-wall system \cite{Adam2}.

In the model discussed in \cite{Adam2}
two central charges --- of the domain wall and domain line types ---
are allowed \cite{CS}. But what we focus on now, is a
different central charge.
The relevant world-volume
central charge in the domain-wall case corresponds to
$CP(1)$  ``lumps."
Although the existence of such states was not explicitly
verified in Ref.~\cite{Adam2} and the corresponding solution not found
due to strong coupling
issues, but the very fact that  composites carrying
this charge do exist in the domain-wall problem is beyond doubt.
Indeed, since the 1/4-BPS (bulk quarter-criticality)
wall junctions (domain lines)
correspond to $CP(1)$ kinks, of which there are two
inequivalent kinds, one
could in principle construct a system with the two domain lines
(on the wall)  joined at a
single point in 1+3 dimensions.
This single point is a ``junction of junctions."
There is no central charge for this localized
junction of junctions in the 1+3 dimensional bulk
theory, but it should nonetheless be a
1/4-BPS state on the wall world volume saturating
both the kink and lump central charges.

\section{Non-Abelian strings in \none SQCD}
\label{6}
\renewcommand{\theequation}{\thesection.\arabic{equation}}
\setcounter{equation}{0}

The IR problems we encounter in  \none SQCD emerging  at $\mu\to\infty$
are quite similar to those discussed in \cite{Yung:1999du,EY}. In these papers
strings on the Higgs branches were studied. In particular, in \cite{EY}
the Abelian strings in \none SQED were
considered\,\footnote{We elaborate on
supertranslational  fermion zero modes   in this theory in Sect.~\ref{7}.}.
This theory has a Higgs branch which can be lifted by embedding
the theory in the deformed \ntwo SQED (\ref{qed}).

In Ref.~\cite{EY} strings at an arbitrary point on the
(lifted) Higgs branch were considered, with both $\langle q\rangle $ and
$\langle\tilde{q}\rangle$ nonvanishing (cf. Eq.~(\ref{qvev}),
where $\langle\tilde{q}\rangle=0$).
In this case the string appears
to be non-BPS. The string solution consists of a ``BPS core'' and a long-range
logarithmic tail of size $\sim 1/m_L$. To take the limit $\mu\to\infty$
one can proceed as follows \cite{Yung:1999du,EY}. Let us consider a string of
a large but
finite length $L$. Then at a very large $r$,
$$r\sim L\,,$$
the problem is no longer
two-dimensional and logarithmic tails are cut off. In
other words, the scale $1/L$ plays the role of the IR cut-off instead of $m_L^{-1}$.
Now one can safely take the limit $\mu\to\infty$.

Let us follow a similar approach to the  problem at hand. Consider the string
of a finite length $L$. Then the scale $1/L$ will play the role of an IR
regularization for the fermion zero modes (\ref{psizero}) and the normalization
integral $I_f$ becomes finite. (Unfortunately, taking the length
of a string to be finite destroys the BPS nature of a string).

Now we can safely take the limit $\mu_2\to\infty$.
The normalization integral for the fermion zero modes (\ref{if}) stays finite,
\beq
I_f \sim g_2^2\, \ln\,{\left(g_2\sqrt{\xi}L\right)}.
\eeq
It still can be absorbed into the definition of the field $\chi_1$.

The world-sheet theory
become non-local
containing  powers of higher derivative corrections, all of the same order.
The non-locality arises because the  string becomes  thick.
Note, that this effect does not affect the string tension.

\section{Abelian strings}
\label{7}
\renewcommand{\theequation}{\thesection.\arabic{equation}}
\setcounter{equation}{0}

In this section we briefly review Abelian BPS strings solutions
and their fermion zero modes in \ntwo SQED
obtained in \cite{VY}, and then elaborate on the issue of
fermion zero modes in the U(1) theory (\ref{qed}),
with broken \ntwo supersymmetry.
In particular, we will focus on the large $\mu$-limit when the
theory (\ref{qed}) reduces to \none SQED.

The Abelian string solution with the
minimal winding number in the model
(\ref{qed}) has the form
\beqn
q(x)
&=&
e^{ i \, \alpha  }\phi(r),
\nonumber\\[4mm]
A_{i}(x)
&=&
- \varepsilon_{ij}\,\frac{x_j}{r^2}\
\left(1-f(r)\right)\,,
\label{abstr}
\eeqn
where $f(r)$ and $\phi(r)$ are profile functions for gauge and scalar
fields, respectively.
These functions satisfy the following first-order equations:
\beqn
&&
r\frac{d}{{d}r}\,\phi (r)-  f(r)\phi (r) = 0\, ,
\nonumber\\[4mm]
&&
-\frac1r\,\frac{ d}{{ d}r} f(r)+\frac{e^2}{4}\,
\left[\left(\phi(r)\right)^2 -\xi\right] =0\, .
\label{abfoe}
\eeqn
The boundary conditions for these functions are
\beq
f(0) = 1\, , \qquad   f(\infty) = 0
\label{abfbc}
\eeq
for the gauge field, while the boundary conditions for  the
squark field are
\beq
\phi (\infty)=\sqrt{\xi}\,,\qquad   \phi (0)=0\, .
\label{abphibc}
\eeq
Equations (\ref{abfoe}) can be solved numerically.
The tension of the  string with the minimal winding is
\beq
T_1=2\pi\,\xi\, .
\label{abten}
\eeq
Note that the string solution does not depend on the deformation
parameter $\mu$,  much in the same way as in the non-Abelian
case. This is because the neutral
scalar field $a$ vanishes on the solution.

Consider first the \ntwo limit $\mu=0$. The string is half-critical,
so 1/2 of supercharges (related to SUSY transformation parameters $\epsilon^{12}$
and $\epsilon^{21}$, see Sect.~\ref{41}) act trivially on the string solution.
The remaining four (real) supercharges
parametrized by $\epsilon^{11}$ and $\epsilon^{22}$
generate four supertranslational fermion zero modes.
They have the form \cite{VY}
\beqn
\bar{\psi}_{\dot{2}}
& = &
-2\sqrt{2}\,\frac{x_1+ix_2}{r^2}\,f\,\phi\,\zeta_2
\, ,
\nonumber\\[3mm]
\bar{\tilde{\psi}}_{\dot{1}}
& = &
2\sqrt{2}\,\frac{x_1-ix_2}{r^2}\,f\,\phi\,\zeta_1 ,
\nonumber\\[5mm]
\bar{\psi}_{\dot{1}}
& = &
0\, , \qquad
\bar{\tilde{\psi}}_{\dot{2}}= 0\, ,
\nonumber\\[4mm]
\lambda^{22}
& = &
ig^2\left(\phi^2-\xi\right)\zeta_1\, ,
\nonumber\\[4mm]
\lambda^{11}
& = &
-ig^2\left(\phi^2-\xi\right)\zeta_2
\,,
\nonumber\\[4mm]
\lambda^{12}
& = & 0
\, ,\qquad  \lambda^{21}= 0\,,
\label{abzmodes}
\eeqn
where the modes proportional to complex Grassmann parameters
$\zeta_1$ and $\zeta_2$ are generated by $\epsilon^{22}$ and
$\epsilon^{11}$  transformations, respectively.

It is quite straightforward to check that these modes satisfy the Dirac equations,
\beqn
&&\frac{i}{e^2} \bar{\pt}\hspace{-0.65em}/\lambda^{f}
+\frac{i}{\sqrt{2}}\,\left(
\bar{\psi} q^f+
\bar{q}^f\bar{\tilde{\psi}}\right)-\mu_2\delta^f_2\bar{\lambda}_2=0,
\nonumber\\[3mm]
&&i\nabla\hspace{-0.65em}/ \bar{\psi}+\frac{i}{\sqrt{2}}\left[\bar{q}_f\lambda^f
+a\,\tilde{\psi}\right]=0,
\nonumber\\[3mm]
&&
i\nabla\hspace{-0.65em}/ \bar{\tilde{\psi}}+
\frac{i}{\sqrt{2}}\left[\lambda_f q^f
+a\,\psi\right]=0
\label{abdirac}
\eeqn
for the U(1) model (\ref{qed}) at $\mu=0$.

Now, we switch on the breaking parameter $\mu$,
$$\mu\neq 0\,.$$
The number of supercharges
in the bulk theory drops to four which means that we have only two supercharges,
associated with the complex parameter $\epsilon^{11}$
acting non-trivially on the string solution. If we apply these supercharges
to the string solution (\ref{abstr}) we generate only half of the modes in
(\ref{abzmodes}) proportional to $\zeta_2$. We get
\beqn
\bar{\psi}_{\dot{2}}
& = &
-2\sqrt{2}\,\frac{x_1+ix_2}{r^2}\,f\,\phi\,\zeta_2\, ,
\nonumber\\[3mm]
\bar{\psi}_{\dot{1}}
& = &
0\, ,
\nonumber\\[4mm]
\lambda^{11}
& = &
-ig^2\left(\phi^2-\xi\right)\zeta_2
\,,
\nonumber\\[4mm]
\lambda^{21}
& = & 0
\, .
\label{abnonemodes}
\eeqn

As in the non-Abelian case the other two zero modes proportional to
$\zeta_1$ do not disappear. They just get modified and can no longer be
obtained by SUSY transformation. We derive them below
by explicitly solving
the Dirac equations (\ref{abdirac}) following the same steps as
in Sects.~\ref{42} and \ref{43}.

First we note that certain  components of the fermion fields are
not generated, namely,
\beq
\bar{\tilde{\psi}}_{\dot{2}}=0, \;\;\;\; \lambda^{12}=0\,.
\label{abdot212}
\eeq
Other components can be parametrized by fermion profile functions
$\lambda_{\pm}(r)$ and  $\psi_{\pm}(r)$ via
\beqn
\lambda^{22}
&=&
\lambda_{+}(r)\,\zeta_1 + \frac{x_1+ix_2}{r}\,
\lambda_{-}(r)\,\bar{\zeta}_1\,,
\nonumber\\[3mm]
\bar{\tilde{\psi}}_{\dot{1}}
&=&
\frac{x_1-ix_2}{r}\,\psi_{+}(r)\,\zeta_1 +
 \psi_{-}(r)\,\bar{\zeta}_1\,.
\label{abfprofile}
\eeqn
The above profile functions  satisfy the following Dirac equations:
\beqn
&&\frac{d}{dr}\psi_{+} +\frac1r\psi_{+}
-\frac1{r} f\psi_{+}+\frac{i}{\sqrt{2}}\,\phi\,
\lambda_{+}=0\,,
\nonumber\\[3mm]
&&-\frac{d}{dr}\lambda_{+}
+i\frac{e^2}{\sqrt{2}}\,\phi\,\psi_{+} +e^2\mu\,\lambda_{-}=0\,,
\nonumber\\[3mm]
&&\frac{d}{dr}\psi_{-}
-\frac1{r}f\psi_{-}+\frac{i}{\sqrt{2}}\,\phi\,\lambda_{-}=0\,,
\nonumber\\[3mm]
&&-\frac{d}{dr}\lambda_{-}-\frac1r\lambda_{-}
+i\frac{e^2}{\sqrt{2}}\,\phi\,\psi_{-} +e^2\mu\,\lambda_{+}=0\,.
\label{abfermeqs}
\eeqn
Parallelizing the derivation in
Sect.~\ref{43} let us consider the large $\mu$-limit,
$$\mu\gg \sqrt{\xi}\,.$$
In this limit we can integrate out
the heavy $\lambda^{22}$ field. In particular,
the first and third equations in (\ref{abfermeqs}) give
\beqn
\lambda_{+}
&=&
\frac{i\sqrt{2}}{\phi}\left[\frac{d}{dr}\psi_{+}
+\frac1r\psi_{+}-\frac{f}{r}\psi_{+}\right]\,,
\nonumber\\[3mm]
\lambda_{-}
&=&
\frac{i\sqrt{2}}{\phi}\left[
\frac{d}{dr}\psi_{-} -\frac{f}{r}\psi_{-}\right]\,.
\label{ablambdapsi}
\eeqn
Neglecting the kinetic terms for $\lambda$ fields in the second and
last equations in (\ref{abfermeqs}) we get
\beqn
&&\frac{d}{dr}\psi_{+} +\frac1r\psi_{-}-\frac{f}{r}\psi_{+}
+\frac{\phi^2}{2\mu}
\,\psi_{-}=0\,,
\nonumber\\[3mm]
&&\frac{d}{dr}\psi_{-} -\frac{f}{r}\psi_{-}+
\frac{\phi^2}{2\mu}\,\psi_{+}=0\,.
\label{abpsieqs}
\eeqn
The large-$r$ behavior in these equations is determined by the mass of the
U(1) gauge multiplet
\beq
\tilde{m}_0=\frac{g}{\sqrt{2}}\,\sqrt{\xi}
\label{abgmass}
\eeq
and the mass of the light chiral multiplet
\beq
\tilde{m}_L=\frac{\xi}{2\mu}\, ,
\label{ablmass}
\eeq
see Sect.~\ref{nabt}. The light mass (\ref{ablmass}) is determined by the
smaller root of the quadratic equation (\ref{abquadeq}). In particular, in
the limit $\mu\to\infty$ it tends to zero. The corresponding massless
states become moduli on the Higgs branch of \none SQED (\ref{noneqed}).

Given this hierarchy of masses we use the same method as in Sect.~\ref{43}
to solve equations (\ref{abpsieqs}).
Consider first the large-$r$ region,
$$r\gg 1/m_0\,.$$
Repeating the same steps
which lead us to Eq.~(\ref{psi}) we get
\beq
\psi_{+}=\tilde{m}_L\sqrt{\xi}\, K_0(\tilde{m}_L r)\,, \qquad
\psi_{-}=-\sqrt{\xi}
 \frac{d}{dr}K_0(\tilde{m}_L r)\, .
\label{abpsi}
\eeq
In particular, at $r\ll 1/\tilde{m}_L$ we have
\beq
\psi_{+}\sim \tilde{m}_L\sqrt{\xi}\, \ln{\frac1{\tilde{m}_L r}}\,, \qquad
\psi_{-}\sim\frac{\sqrt{\xi}}{r}\, .
\label{abpsizero}
\eeq
Passing to the intermediate region of $r$,
$$r\le 1/\tilde{m}_0\,,$$
we now obtain
\beq
\psi_{+}=\tilde{m}_L \ln{(\tilde{m}_0/\tilde{m}_L)}
\,\phi\, ,\qquad
\psi_{-}=\frac{1}{r}\,\phi\, .
\label{absdpsi}
\eeq

Equation (\ref{abpsi}) shows
that the supertranslational fermion zero modes  of the Abelian string
in the model (\ref{qed}) acquire long-range tails too. In particular, in the
limit $\mu\to\infty$ they become logarithmically non-normalizable.
Still at
any finite $\mu$ we can absorb the normalization integral into the
definition of the two-dimensional fermion fields $\zeta_1$,
exactly in the same way this was done for the
superorientational modes in Sect.~\ref{5}. This leads us to the
following effective
theory on the world sheet of the Abelian string:
\beqn
S_{1+1}
&=&
2\pi\xi \int d t d z \left\{\frac12
\left(\pt_k x_{0i}
\right)^2+
\frac12 \, \bar{\zeta}_1 \, i
\left(\pt_0-i\pt_3
\right)\, \zeta_1
\right.
\nonumber\\[3mm]
&+& \left.
\frac12 \, \bar{\zeta}_2 \, i
\left(\pt_0+i\pt_3
\right)\, \zeta_2
\right\}\, ,
\label{ntwofree}
\eeqn
where $x_{0i}$ ($i=1,2$) denote the coordinates of the string position
in $(1,2)$-plane.

This is a free theory with two real bosonic and four fermionic fields
of $t,\,z$.
Counting  the number of degrees of freedom we observe the
enhanced \ntwo
supersymmetry in two dimensions (four supercharges):
the fields at hand form a supermultiplet of \ntwo.

We see that the
phenomenon of the enhanced world-sheet supersymmetry is quite general
and occurs both for Abelian and non-Abelian strings. It can be traced back to
strings in \ntwo supersymmetric bulk theory from which our strings
are descendants.

The-two dimensional theory (\ref{ntwofree}) is a trivial free-field
theory and it
does not generate its own scale. Therefore, we cannot estimate the
critical value of $\mu$ when the enhanced \ntwo supersymmetry breaks down
to \none in this case. The theory (\ref{ntwofree}) is a low-energy
effective theory which describes the string at small energies $E$,
$E\ll m_L$. At larger energies higher derivative corrections to
(\ref{ntwofree}) become important. The higher derivative sector does not
respect \ntwo supersymmetry and at large energies
supersymmetry breaking effects take over. As we increase $\mu$,
the region of
validity of (\ref{ntwofree}) becomes exceedingly narrower. In the
\none SQED limit of $\mu\to\infty$ the string becomes
thick and the effective theory on the string world sheet becomes
non-local.
It is worth stressing again that this happens due to the presence of
the Higgs
branch in \none SQED.

There is one more thing we must emphasize.
The translational sector
of the U(1) gauge theory (\ref{qed})
is discussed  in this section just for the sake
of simplicity.
The generalization to the translational sector of
the non-Abelian string
in theory (\ref{model}) is
absolutely straightforward.
We get the same results for the translational sector of
the non-Abelian string.

\section{Conclusions}
\label{8}
\renewcommand{\theequation}{\thesection.\arabic{equation}}
\setcounter{equation}{0}

This concluding section could have been entitled
{\sl ``How extended supersymmetry dynamically emerges from
K\"ahlerian geometry."} After the phenomenon
is identified, it seems to be rather trivial and transparent.
Indeed, if we start from a bulk theory
with\,\footnote{Although $\nu=4$ in the 4D cases under
consideration we would like to stick to a more general formulation.}
$\nu$ supercharges and obtain half-critical solitons
with a nontrivial moduli space, a linear realization of $\nu/2$
supercharges in the low-energy world-sheet theory of moduli
is guaranteed. If, in addition, the geometry of the moduli
space is K\"ahlerian, and the numbers of the boson and fermion
zero modes appropriately match, $\nu/2$ extra ``supernumerary" supercharges
emerge with necessity. Apparently this is not a rare
occurrence, since we encounter one and the same situation, enhancement
of supersymmetry,
in two most widely discussed problems --- domain walls in \none SQCD
with $N_f=N_c$ \cite{Adam2}, and in the current problem of non-Abelian
strings. It is worth stressing, however, that
the reasons lying behind enhancement
of supersymmetry in these two problems are not quite the same, as
was explained in Sect.~\ref{intro}.

We also observe ``supersymmetry emergence"
for the flux-tube junctions (confined monopoles):
our kink--monopole is half-critical considered
from the standpoint of the world-sheet $CP(1)$ model
(i.e. two supercharges conserved),
while in the bulk \none theory there is
no monopole central charge at all.
A similar phenomenon was also noted in Ref.~\cite{Adam2}.

\vspace{2mm}

A number of interesting questions
remains unanswered or not answered in full.
Let us list some of them.

(i) In Sect. III.B3 of Ref.~\cite{Adam2} it was shown
that a mass deformation
removing the continuous moduli space
of the world-volume theory
leaves the enhanced \ntwo supersymmetry
intact, at least for small mass deformations.
The lifting of the moduli space occurred through a generation
of a Killing vector potential. More precisely, it
was verified that, at leading order in the unequal mass deformation,
the effect of the mass deformation reduced to a potential which is the norm-squared of
a U(1) Killing vector on $CP(1)$ (the so-called real mass deformation),
see \cite{Alvarez}.
Such a potential preserves \ntwo as
it maintains the complex structure. It was unclear what symmetry  ensures this
form for the potential. It was also unclear whether this particular form
holds beyond the leading order in the deformation.

It would be extremely interesting to explore whether or not a similar
structure persists in the flux-tube case.

(ii) It seems imperative to understand the necessary (rather than just
sufficient)  conditions for supersymmetry enhancement more precisely.
In the context of this question
it would be nice to find a symmetry argument which 
would explain why turning on a
(finite)  adjoint mass has no impact (up to field rescalings) on the
flux-tube world-sheet theory.

(iii) Another interesting question is:
what happens with our ``kink--monopole" state in \none theory
when we vary parameters moving towards weaker confinement?
In other words a challenging and illuminating
problem is:   what happens
when two scales in Eq.~(\ref{weakcoupling}) are of the same order?
We are not aware of any discussion
of this regime in the literature.

(iv) The issue of supersymmetry emergence seems intriguing.
Is it promising from the standpoint of applications?

\vspace{2mm}

We hope to return to the above issues elsewhere.

\newpage

\section*{Acknowledgments}

We are very grateful to Adam Ritz  for thorough
discussions and communications, and careful reading
of a preliminary version of the present paper
which resulted in a number of valuable remarks.
We would like to thank
Nathan Seiberg for posing stimulating questions.

The work  of M.S. was supported in part by DOE
grant DE-FG02-94ER408.
The work of A.Y. was supported in part by the RFBF grant
No.~05-02-17360 and by Theoretical Physics Institute at the
University of Minnesota.

\end{document}